\documentclass[prd,preprint,superscriptaddress,preprintnumbers,eqsecnum,showpacs,nofootinbib,nobibnotes]{revtex4-1}
\usepackage{amsfonts,amsmath,amssymb,bm}
\usepackage[table]{xcolor}
\usepackage{graphicx} 
\usepackage{mathtools}
\usepackage{boondox-cal}
\usepackage{slashed}
\usepackage{xcolor}

\usepackage[utf8]{inputenc}
\usepackage{slashed}

\newcommand{\be}{\begin{equation}}
\newcommand{\bea}{\begin{eqnarray}}
\newcommand{\ba}{\begin{align}}
\newcommand{\ee}{\end{equation}}
\newcommand{\eea}{\end{eqnarray}}
\newcommand{\ea}{\end{align}}

\newcommand{\cntrm}{\overline{\Gamma}}
\newcommand{\finite}{F}
\newcommand{\ccntrm}{\gamma}
\newcommand{\ffinite}{f}

\definecolor{zero2}{rgb}{0.88,0.88,.88}

\def\1eq#1{Eq.~(\ref{#1})}

\def\2eqs#1#2{Eqs.~(\ref{#1}) and~(\ref{#2})}
\def\3eqs#1#2#3{Eqs.~(\ref{#1}),~(\ref{#2}) and~(\ref{#3})}
\def\4eqs#1#2#3#4{Eqs.~(\ref{#1}),~(\ref{#2}),~(\ref{#3}) and~(\ref{#4})}
\def\noeq#1{(\ref{#1})}

\def\s#1{{\scriptscriptstyle #1}}

\def\G{\Gamma}

\def\Gwt{\widehat{\Gamma}}

\def\s{\mathcal{s}}

\def\hphi0{{\hat\phi}_0}

\def\d{\!\mathrm{d}^4x\,}

\def\dP{\!\mathrm{d}^D p}
\def\dt{\!\mathrm{d} t\,}

\def\cfxt#1{\overline{\theta}_{#1}}
\def\cfct#1{\overline{\rho}_{#1}}
\def\cfgi#1{\overline{\lambda}_{#1}}
\def\cfps#1{\overline{\vartheta}_{#1}}
\def\ff#1#2{\gamma^{#1}_{#2}}

\makeatletter
\def\user@resume{resume}
\def\user@intermezzo{intermezzo}
\newcounter{previousequation}
\newcounter{lastsubequation}
\newcounter{savedparentequation}
\renewenvironment{subequations}[1][]{%
      \def\user@decides{#1}%
      \setcounter{previousequation}{\value{equation}}%
      \ifx\user@decides\user@resume 
           \setcounter{equation}{\value{savedparentequation}}%
      \else  
      \ifx\user@decides\user@intermezzo
           \refstepcounter{equation}%
      \else
           \setcounter{lastsubequation}{0}%
           \refstepcounter{equation}%
      \fi\fi
      \protected@edef\theHparentequation{%
          \@ifundefined {theHequation}\theequation \theHequation}%
      \protected@edef\theparentequation{\theequation}%
      \setcounter{parentequation}{\value{equation}}%
      \ifx\user@decides\user@resume 
           \setcounter{equation}{\value{lastsubequation}}%
         \else
           \setcounter{equation}{0}%
      \fi
      \def\theequation  {\theparentequation  \alph{equation}}%
      \def\theHequation {\theHparentequation \alph{equation}}%
      \ignorespaces
}{%
  \ifx\user@decides\user@resume
       \setcounter{lastsubequation}{\value{equation}}%
       \setcounter{equation}{\value{previousequation}}%
  \else
  \ifx\user@decides\user@intermezzo
       \setcounter{equation}{\value{parentequation}}%
  \else
       \setcounter{lastsubequation}{\value{equation}}%
       \setcounter{savedparentequation}{\value{parentequation}}%
       \setcounter{equation}{\value{parentequation}}%
  \fi\fi
  \ignorespacesafterend
}
\def\CT@@do@color{%
	\global\let\CT@do@color\relax
		\@tempdima\wd\z@
		\advance\@tempdima\@tempdimb
		\advance\@tempdima\@tempdimc
		\advance\@tempdimb\tabcolsep
		\advance\@tempdimc\tabcolsep
		\advance\@tempdima1.5\tabcolsep
	\kern-1.5\@tempdimb
	\leaders\vrule
	\hskip\@tempdima\@plus  1fill
	\kern-1.5\@tempdimc
	\hskip-\wd\z@ \@plus -1fill }
\makeatother

\begin{document}

\title{%One-loop
Renormalizable Extension of the Abelian Higgs-Kibble
Model with a dimension 6 operator
}

\date{April 19, 2022}

\author{D. Binosi}
\email{binosi@ectstar.eu}
\affiliation{European Centre for Theoretical Studies in Nuclear Physics
and Related Areas (ECT*) and Fondazione Bruno Kessler, Villa Tambosi, Strada delle Tabarelle 286, I-38123 Villazzano (TN), Italy}
\author{A. Quadri}
\email{andrea.quadri@mi.infn.it}
\affiliation{INFN, Sezione di Milano, via Celoria 16, I-20133 Milano, Italy}

\begin{abstract}
\noindent
A deformation of the Abelian Higgs Kibble model induced by a dimension 6 derivative operator is studied. A novel differential equation is established fixing the dependence of the vertex functional on the coupling $z$ of the dim.6 operator
in terms of amplitudes at $z=0$ (those of the 
power-counting renormalizable Higgs-Kibble model).
The latter equation holds in a formalism where the 
physical mode is described by a gauge-invariant field.
The functional identities of the theory in this formalism are studied. In particular we show that the Slavnov-Taylor identities
separately hold true at each order in the number of internal
propagators of the gauge-invariant scalar. 
Despite being non-power-counting renormalizable,
the model at $z\neq 0$ depends on a finite number of physical parameters.
\end{abstract}


\maketitle

\section{Introduction}

The experimental program at High-Luminosity and High-Energy LHC~\cite{Cepeda:2019klc} will provide a unique opportunity to explore the Higgs sector of the electroweak theory, thus elucidating the electroweak spontaneous symmetry breaking mechanism (EWSSB). To be sure, while the Standard Model's requirement of gauge-invariance and power-counting renormalizability uniquely predicts the quartic Higgs potential as the source of EWSSB, many other alternatives exist in Beyond the Standard Model (BSM) theories where either new particles are introduced while preserving power-counting renormalizability ({\it e.g.}, in the case of the Two-Higgs Doublet Model or the Minimal Supersymmetric Standard Model) or additional power-counting violating interactions are switched on in the spirit of effective field theories~\cite{Buchmuller:1985jz,Alonso:2013hga,Brivio:2017vri}.

In the latter case, the higher dimensional operators introduced must fulfil the relevant symmetries of the specific theory considered, {\it e.g.}, gauge symmetry, Lorentz covariance and further possible discrete symmetries. The couplings of such operators are otherwise unconstrained additional physical parameters that must be fixed by suitable normalization conditions. Their number increases order by order in the loop expansion, since more and more ultaviolet (UV) divergences arise in those effective models, as a consequence of the lack of power-counting renormalizability.

It sometimes happens that in a non power-counting renormalizable theory some of the coefficients of the higher dimensional operators  can be reduced to a smaller number of independent ones. This is what happens for instance in the reduction of couplings approach~\cite{Heinemeyer:2019vbc}, since reduced couplings are functions of a primary one, satisfying a set of differential equations compatible with the renormalization group 
%(RG) 
flow. Additionally, it is well-known that the appropriate choice of field coordinates can greatly simplify the task of identifying the independent couplings of the model. For instance, one can obtain an equivalent theory with non-renormalizable couplings from a power-counting renormalizable one by applying an invertible non-linear field transformation. Provided that a suitable prescription is adopted when quantizing the former theory, physical observables in the two models do not differ, as stated by the so-called Equivalence Theorem~\cite{Kallosh:1972ap,Kamefuchi:1961sb,Ferrari:2002kz}

As a preliminary study for the treatment of the full electroweak model, in this paper we consider a specific extension of the power-counting renormalizable Abelian Higgs-Kibble theory  quantized in the simplest linear representation of the physical Higgs scalar via a gauge-invariant variable $X_2$, reducing on-shell to the gauge-invariant combination~\cite{Binosi:2017ubk,Binosi:2019olm,Binosi:2019nwz,Binosi:2020unh}\begin{align}
    X_2 \sim \frac{1}{v} \left ( \phi^\dagger \phi - \frac{v^2}{2} \right ).
    \label{x2def}
\end{align}
Within this framework, we find that there is a unique deformation of the power-counting renormalizable model that preserves at the quantum level the defining  functional identities of the Abelian Higgs-Kibble model; more specifically, this deformation is induced by a modified kinetic term
\begin{align}
	-\frac{z}{2} X_2 \square X_2,
\end{align}
which corresponds, after the identification~\noeq{x2def}, to the dimension 6 derivative operator
\begin{align}
    -\frac{z}{2v^2}
    \left ( \phi^\dagger \phi - \frac{v^2}{2} \right ) \square
    \left ( \phi^\dagger \phi - \frac{v^2}{2} \right ).
        \label{d6.op}
\end{align}
At $z=0$ one recovers the power-counting renormalizable Higgs-Kibble model, whereas at $z\neq 0$ power-counting renormalizability is lost; nevertheless, one can write a $z$-differential equation connecting the dependence of the one-particle irreducible (1-PI) amplitudes on the coefficient $z$ and the mass $M$ of the physical scalar. Then, under some reasonable assumptions on the boundary conditions a unique solution exists for the 1-PI vertex functional (the generating functional of the 1-PI amplitudes) of the deformed theory, fulfilling the $z$-differential equation and all the remaining symmetries of the theory. Such vertex functional at $z\neq 0$ is constructed out  of the Feynman amplitudes of the power-counting renormalizable theory at $z=0$; most notably, each subsector of the theory at $z=0$, labelled by the number $\ell$ of internal lines of the physical scalar $X_2$ propagating inside loops, can be lifted in a unique way to the corresponding subsector at $z\neq 0$.

The consistency of the solutions to the $z$-differential equation we will construct comes from the property that,  at variance with the conventional representation of the scalar Higgs field~$\phi$, the $\ell$-subsectors of the theory are separately Slavnov-Taylor invariant. Hence, despite not being power-counting renormalizable, the theory at $z \neq 0$ can still be defined in terms of a finite number of physical parameters (those of the power-counting renormalizable theory and $z$). Thus, once extended to the electroweak gauge group, the distinct phenomenological implications of such a theory can be identified and tested against the available experimental results~\cite{Binosi:2022prp}.
  
The present paper is organized as follows. In Sect.~\ref{sec.intro} we introduce the model and set our notations. Next, in Sect.~\ref{sec.zeq} we derive the differential equation controlling the dependence of the theory on the deformation parameter $z$. Renormalization of the $z$-differential equation and the ensuing constraints on the 1-PI amplitudes are derived in Sect.~\ref{sec.constraints}; the corresponding Slavnov-Taylor (ST) identity is then studied in Sect.~\ref{sec.sti}, together with its decomposition in a tower of relations among 1-PI Green's functions at fixed order in the number of internal $X_2$-lines. %Consequences of the latter decomposition on the gauge-invariant coupling constants are studied in Sect.~\ref{sec.cpls}, whereas 
The compatibility of the $z$-differential equation with different renormalization schemes is studied in Sect.~\ref{sec:otherrens} and our conclusions and outlook are finally presented in Sect.~\ref{sec:conclusions}.

\section{Extension of the Abelian Higgs-Kibble model}\label{sec.intro}

We will consider the Abelian Higgs-Kibble (HK) model~\cite{Becchi:1974md,Becchi:1974xu} extended  with the dimension 6 operator in Eq.(\ref{d6.op})
%\begin{align}
%    - \frac{z}{2 v^2} \left ( \phi^\dagger \phi - \frac{v^2}{2} \right ) \square 
%    \left ( \phi^\dagger \phi - \frac{v^2}{2} \right )
%\end{align}
as a useful playground in view of the treatment of the full $\rm SU(2) \times U(1)$ electroweak theory with mass generation {\em \`a la} Higgs. 
In \1eq{d6.op} $\phi$ is the complex Higgs field
\begin{align}
	\phi = \frac{1}{\sqrt{2}} ( \sigma + v + i \chi),
\end{align} 
with $v$  its  vacuum expectation value; the field $\sigma$ describes the physical scalar mode while  $\chi$ is the Goldstone field. Finally, $z$ represents the parameter controlling the non-power-counting renormalizable deformation induced by the dimension 6 operator in \1eq{d6.op}.

We use a gauge-invariant coordinate in order to describe the physical scalar model, in accord with the formalism of~\cite{Quadri:2006hr,Quadri:2016wwl,Binosi:2017ubk}. It is only within such an approach that one can obtain the differential equation constraining the dependence of the 1-PI amplitudes on $z$ and, with it, all the results discussed in the following sections. To this end, one introduces the field $X_2$ together with a Lagrange multiplier $X_1$ to obtain the vertex functional reported in~\1eq{tree.level}~\cite{Quadri:2006hr,Quadri:2016wwl,Binosi:2017ubk}; the auxiliary $X_i$ fields are then such that when going on-shell one obtains
\begin{align}
    X_2 \sim \frac{1}{v} \left ( \phi^\dagger \phi - \frac{v^2}{2} \right ),
\end{align}
so that under this replacement the term $-\frac{z}{2} X_2 \square X_2$ in \1eq{tree.level} reduces to the one in \1eq{d6.op} and we get back the Abelian HK model supplemented by the latter dimension 6 operator.

The main advantage of the $X$-representation of the physical scalar mode is two-fold. On the one hand, the full dependence on the additional parameter $z$ is contained in the quadratic part of the classical vertex functional, so that in the perturbative expansion using the mass eigenstate basis (leading to the diagonal propagators) the coupling $z$ will enter in the $X_2$-propagator but not in the  interaction vertices, contrary to what happens in the conventional $\phi$-representation of \1eq{d6.op}. On the other hand, since $X_2$ is gauge-invariant, the projection of the Slavnov-Taylor (ST) identity valid for the 1-PI amplitudes yields an additional set of relations which isolate separately invariant subsectors of the theory according to the number of internal $X_2$-lines (with the lowest order reproducing the St\"uckelberg theory, namely no internal $X_2$-lines).

\section{Differential Equation for $z$}\label{sec.zeq}

In the mass eigenstate basis the dependence on the parameter $z$ only arises via the $X_2$-propagator, see \1eq{tree.level}; indeed, an $X_2$ line circulating inside a general $n$-loop diagram will be characterized by a propagator $\Delta_{X_2X_2}$ given by, see \1eq{app.prop.1}
\begin{align}
    \Delta_{X_2X_2}(k^2,M^2) = \frac{i}{(1+z)k^2-M^2}.
\end{align}
Introducing then the differential operator
\begin{align}
    {\cal D}_z^{M^2}=(1+z)\partial_z+M^2\partial_{M^2},
\end{align}
one finds that $\Delta_{X_2X_2}$ is an eigenvector of ${\cal D}_z^{M^2}$ with eigenvalue -1:
\begin{align}
    {\cal D}_z^{M^2}\Delta_{X_2X_2}(k^2,M^2)=-\Delta_{X_2X_2}(k^2,M^2).
\end{align}

Next, let us collectively denote with $\Phi$ the set of fields and external sources of the theory, and let us indicate with $p_i$ (with $i=1,\dots,r$) the external momenta, with $\Phi_i=\Phi(p_i)$ and $p_r=-\sum_1^{r-1} p_i$; in this way a $n$-loop 1-PI Green's function $ \G^{(n)}_{\Phi_1\cdots\Phi_r} $ with $r$ $\Phi_i$ insertions can be decomposed as the sum of all diagrams with (amputated) external legs $\Phi_1\cdots\Phi_r$ with zero, one, two,..., $\ell$ internal $X_2$-propagators, {\it i.e.},
\begin{align}
    \G^{(n)}_{\Phi_1\cdots\Phi_r} = 
    \sum_{\ell\geq 0}\G^{(n;\ell)}_{\Phi_1\cdots \Phi_r} .
    \label{1pi.exp}
\end{align}
Then, clearly,
\begin{align}
	{\cal D}^{M^2}_z \G^{(n;\ell)}_{\Phi_1\cdots\Phi_r} &= -
    \ell\G^{(n;\ell)}_{\Phi_1\cdots \Phi_r};& &\Longrightarrow& 
    {\cal D}^{M^2}_z \G^{(n)}_{\Phi_1\cdots\Phi_r} = -
    \sum_{\ell\geq 0}\ell\G^{(n;\ell)}_{\Phi_1\cdots \Phi_r}.
    \label{eq.1pi.exp}
\end{align}
The most general solution of this equation reads (indicating explicitly only the dependence on the parameters $z$ and $M^2$)
\begin{align}
	\G^{(n;\ell)}_{\Phi_1\cdots\Phi_r}(z,M^2)=\frac1{(1+z)^\ell}\G^{(n;\ell)}_{\Phi_1\cdots\Phi_r}(0,M^2/1+z).
	\label{genstr}
\end{align}
Thus, amplitudes at $z\neq0$ in each $\ell$-sector are obtained from those at $z=0$ by dividing them by the $(1+z)^{\ell}$ factor and rescaling by $(1+z)$ the square of the Higgs mass $M^2$.
 
This constitutes already a powerful result. For example consider the theory's $\beta$ functions
\begin{align}
    \beta_i  = (4 \pi)^2 \frac{d}{d \log \mu^2} \overline{C}_i = 
    (4 \pi)^2 \overline{C}_i,
\end{align}
where $\overline{C}_i$ is the residue of the pole in $1/\epsilon$ of the  coefficient $C_i$ of the corresponding one-loop ST invariant. Then the $z$-dependence is recovered by making the replacement $M^2 \rightarrow \frac{M^2}{1+z}$ and the rescaling $1/(1+z)^\ell$ in the $z=0$ coefficients derived in~\cite{Binosi:2019nwz} (with the corresponding results reported for convenience in Appendix~\ref{app.invs}). Additionally, the one-loop $\beta$-functions will also inherit the grading according to the number of internal $X_2$-lines via the grading of the $C_i$ coefficients.

Now, since the combinatorial coefficient 
 in the r.h.s. of \1eq{eq.1pi.exp} depends on the number of internal $X_2$-lines, the r.h.s. cannot be expressed in closed form as a function of $\G^{(n)}_{\Phi_1\cdots\Phi_r}$; yet, it is possible to derive a differential equation for an extension of the full vertex functional $\G$ that takes appropriately into account these factors. To this end, let us  define a modified 1-PI Green's function which depends on an auxiliary parameter $t$ in such a way that 
\begin{align}
    \G^{(n)}_{\Phi_1\cdots\Phi_r}(t) &=
     \G^{(n;0)}_{\Phi_1\cdots\Phi_r} +
     \sum_{\ell\geq 1} t^{\ell-1} \G^{(n;\ell)}_{\Phi_1\cdots\Phi_r};&
     \G^{(n)}_{\Phi_1\cdots\Phi_r}(1) &=
     \G^{(n)}_{\Phi_1\cdots\Phi_r}. 
    \label{1pi.exp.t}
\end{align}
Then,  applying the differential operator ${\cal D}^{M^2}_z$ on the left-hand side of the equation above, using the fact that ${\cal D}^{M^2}_z \G^{(n;0)}_{\Phi_1\cdots\Phi_r} =0 $, and  integrating over $t$ between $0$ and $1$, we find 
\begin{align}
{\cal D}^{M^2}_z \int_0^1 \dt  \G^{(n)}_{\Phi_1\cdots\Phi_r}(t) & =  \sum_{\ell\geq 1} 
\int_0^1 \dt ~ t^{\ell-1} ~{\cal D}^{M^2}_z \G^{(n;\ell)}_{\Phi_1\cdots\Phi_r}
=-\sum_{\ell\geq 1}\int_0^1 \dt\, \ell\, t^{\ell-1}\G^{(n;\ell)}_{\Phi_1\cdots\Phi_r}
\nonumber \\
&= - \sum_{\ell\geq 1}  \G^{(n;\ell)}_{\Phi_1\cdots\Phi_r}
= - \G^{(n)}_{\Phi_1\cdots\Phi_r} +
\G^{(n;0)}_{\Phi_1\cdots\Phi_r} .
\end{align} 
where in the last step we have used the definition~\noeq{1pi.exp.t}. Collecting finally the Green's functions in the $t$-dependent generating functional 
\begin{align}
    \G(t) = \sum_{n,\Phi,r} \int \dP_1 \dots \dP_r ~
    w_{\Phi_1\cdots\Phi_r} \G^{(n)}_{\Phi_1\cdots\Phi_r}(t) ~ \Phi_1\cdots \Phi_r ,
    \label{Gammat}
\end{align}
where $ w_{\Phi_1\cdots\Phi_r}$ are suitable combinatorial weights ({\it e.g.}, if all $\Phi$'s are the same field $w_{\Phi_1\cdots\Phi_r} = 1/r!$),  we obtain the announced $z$-differential equation:
\begin{align}
\int_0^1 \dt {\cal D}^{M^2}_z \G(t) = 
- \G(1) + \G_0.
\label{z.eq}
\end{align}
In the equation above, $\G_0$  represents the generating functional of the 1-PI amplitudes without internal $X_2$-lines ($\ell=0$), 
\begin{align}
\G_0 = \sum_{n,\Phi,r} \int \dP_1 \dots \dP_r ~
      w_{\Phi_1\cdots\Phi_r} \G^{(n;0)}_{\Phi_1\cdots\Phi_r}\Phi_1\cdots \Phi_r ,
\end{align}
which coincides with that of the St\"uckelberg sector of the theory; finally, $\G(1)$ is the vertex functional of the complete theory we are interested in.

%\subsection{Small-$z$ expansion}

Now let us assume $z$ to be small (notice that in the SM this assumption would be natural as $z$ represents the parameter controlling the SM non-power-counting renormalizable deformation). This will allow in turn to expand the functional $\G(t)$ in powers of $z$:
\begin{align}
    \G(t) = \sum_k z^k \G_{[k]}(t),
    \label{exp.G}
\end{align}
with $\G_{[k]}(t)$ independent of $z$. Notice that $\G_{[0]}$ is the power-counting renormalizable theory at $z=0$; and that, since~Eq.\noeq{exp.G} holds true to all orders in the loop expansion, each $\G_{[k]}(t)$ receives contributions from all the different loop orders.

Plugging \1eq{exp.G} into \1eq{z.eq} and projecting at different orders in $z$ we find:
\begin{subequations}
\begin{align}
    &{\cal O}(1):& &\int_0^1 \dt
    \left[ \G_{[1]} (t) + M^2\partial_{M^2} \G_{[0]} (t) \right] = - \G_{[0]}(1) + \G_0;
    \label{ts.1}\\
    &{\cal O}(z):& & \int_0^1 \dt
    \left[  2 \G_{[2]} (t) +
    \G_{[1]}(t) + 
    M^2\partial_{M^2} \G_{[1]}(t) \right] = - \G_{[1]}(1); \label{ts.2}\\
    &{\cal O}(z^2):& & \int_0^1 \dt
    \left[ 3 \G_{[3]} (t) +
    2\G_{[2]}(t) + 
    M^2\partial_{M^2} \G_{[2]} (t) \right] = - \G_{[2]}(1);\label{ts.3}\\
    &\hspace{0.4cm}\vdots& &\hspace{5.cm}\vdots&\nonumber \\
    &{\cal O}(z^k):& &\int_0^1 \dt
    \left[ (k+1) \G_{[k+1]} (t) +
    k\G_{[k]}(t) + 
     M^2\partial_{M^2} \G_{[k]} (t)\right] = - \G_{[k]}(1).\label{ts.k}
\end{align}
\label{ts}	
\end{subequations}

As before, each functional $\G_{[k]}(t)$ can be expanded according to the double grading with respect to the loop number $n$ and the $\ell$-sector; %number of internal $X_2$-lines $\ell$; 
and the tower of Eqs.~(\ref{ts}) yields a set of relations among the amplitudes of the non-renormalizable theory at $z\neq 0$. At the lowest order they depend on the amplitudes of the power-counting renormalizable theory at $z=0$ through the term $\G_{[0]}(1)$ in the r.h.s. of Eq.(\ref{ts.1}) and on those of the St\"uckelberg subsector $\G_0$. Explicit examples are provided in the next subsection.

\subsection{Explicit checks}

Let us carry out some explicit checks that the $z$-differential equation is indeed satisfied by the regularized one-loop 1-PI amplitudes. Constraints on the finite counter-terms arising from the equation will be discussed later on, in Sect.~\ref{sec.constraints}.

\subsubsection{Tadpoles}

Consider the one-loop ($n=1$) tadpoles of the field $\sigma$ and the anti-field $\bar c^*$. From the vertex functional~\noeq{tree.level} one sees that both tadpoles decompose at the one-loop order as the sum of a part with no internal $X_2$-lines and a one-diagram with one ($\ell=1$) internal $X_2$-line. Thus,~\1eq{1pi.exp.t} reads ($\Phi=\sigma,\bar c^*$)
\begin{align}
	\Gamma_\Phi^{(1)}(t)=\Gamma_\Phi^{(1)}=\Gamma^{(1;0)}_\Phi+\Gamma_\Phi^{(1;1)},
\end{align}
which must be replaced everywhere in the entire tower of Eqs.~(\ref{ts}), that is one has
\begin{align}
	\Gamma_{[\cdots]}(t)\to\Gamma^{(1)}_{[\cdots]\Phi}=\Gamma^{(1;0)}_{[\cdots]\Phi}+\Gamma_{[\cdots]\Phi}^{(1;1)}
\end{align}
Considering only the UV divergent parts (denoted with a bar), an explicit calculation gives (in the Feynman gauge $\xi=1$) 
\begin{subequations}
\begin{align}
	\overline{\G}^{(1;0)}_{\sigma} &= \frac{1}{16 \pi^2} \frac{M_A^2}{v \epsilon} (m^2 + 6 M_A^2); &
    \overline{\G}^{(1;1)}_{\sigma} &= \frac{1}{16 \pi^2 } \frac{M^2}{(1+z)^3 v\epsilon} \left[ m^2 (1+z) + 2 M^2 \right], \\
    \overline{\G}^{(1;0)}_{\bar c^*} &= - \frac{1}{16 \pi^2}\frac{M_A^2}{\epsilon}; &
    \overline{\G}^{(1;1)}_{\bar c^*} &= - \frac{1}{16 \pi^2} \frac{M^2}{(1+z)^2\epsilon}. 
\end{align}
\end{subequations}
Expanding around $z=0$ yields, up to ${\cal O}(z^2)$
\begin{subequations}
	\begin{align}
		\overline{\G}^{(1;1)}_{[0]\sigma} &=\frac1{16\pi^2}\frac{M^2}{v\epsilon}\left(m^2+2M^2\right);&
		\overline{\G}^{(1;1)}_{[0]\bar c^*} &=-\frac1{16\pi^2}\frac{M^2}{\epsilon},\\
		\overline{\G}^{(1;1)}_{[1]\sigma} &=-\frac1{8\pi^2}\frac{M^2}{v\epsilon}\left(m^2+3M^2\right);&
		\overline{\G}^{(1;1)}_{[1]\bar c^*} &=\frac1{8\pi^2}\frac{M^2}{\epsilon},\\
		\overline{\G}^{(1;1)}_{[2]\sigma} &=\frac3{16\pi^2}\frac{M^2}{v\epsilon}\left(m^2+4M^2\right);&
		\overline{\G}^{(1;1)}_{[2]\bar c^*} &=-\frac3{8\pi^2}\frac{M^2}{\epsilon}.
	\end{align}
\end{subequations}
Then it is immediate to show
\begin{subequations}
\begin{align}
	\overline{\G}^{(1;1)}_{[1]\Phi}+M^2\partial_{M^2}\overline{\G}^{(1;1)}_{[0]\Phi}&=-\overline{\G}^{(1;1)}_{[0]\Phi},\\
	2\overline{\G}^{(1;1)}_{[2]\Phi}+\overline{\G}^{(1;1)}_{[1]\Phi}+M^2\partial_{M^2}\overline{\G}^{(1;1)}_{[1]\Phi}&=-\overline{\G}^{(1;1)}_{[1]\Phi}.
\end{align} 
\end{subequations}

\subsubsection{2-point functions}

More interesting is the case of the 2-point functions, as in this case an explicit $t$-dependence can arise from diagrams involving two internal $X_2$-lines:
\begin{align}
	\Gamma_{\Phi_1\Phi_2}^{(1)}(t)=\Gamma_{\Phi_1\Phi_2}^{(1;0)}+\Gamma_{\Phi_1\Phi_2}^{(1;1)}+t\Gamma_{\Phi_1\Phi_2}^{(1;2)}.
\end{align} 
with the corresponding replacement in Eqs.~(\ref{ts})
\begin{align}
	\Gamma_{[\cdots]}(t)\to\Gamma^{(1)}_{[\cdots]\Phi_1\Phi_2}(t)=\Gamma^{(1;0)}_{[\cdots]\Phi_1\Phi_2}+\Gamma_{[\cdots]\Phi_1\Phi_2}^{(1;1)}+t\,\Gamma_{[\cdots]\Phi_1\Phi_2}^{(1;2)}.
\end{align}
This is the case when $\Phi_1\Phi_2=\bar c^*\bar c^*,\ \bar c^*\sigma$ and $\sigma\sigma$; instead, the 2-point functions $\Phi_1\Phi_2=\chi\chi,\ \chi A_\mu$ and $A_\mu A_\nu$ do not present such diagrams (and therefore the check of the $z$-differential equation proceeds as in the tadpole case).  

Consider then the $\bar c^*\bar c^*$ function for which an explicit calculation yields
\begin{align}
\overline{\G}^{(1;0)}_{\bar c^* \bar c^*} &= \frac{1}{16 \pi^2} \frac{1}{\epsilon};&
\overline{\G}^{(1;1)}_{\bar c^* \bar c^*} &= 0;&
\overline{\G}^{(1;2)}_{\bar c^* \bar c^*} &= \frac{1}{16 \pi^2} \frac{1}{(1+z)^2} \frac{1}{\epsilon}.
\label{uv.2barc}
\end{align}
Then, noticing that there is no dependence on $M^2$ and that the $n=1$ $\ell=1$ term is convergent, \1eq{ts.1} reads
\begin{align}
	\int_0^1 \dt \,  t\, \overline{\G}^{(1;2)}_{[1]\bar c^* \bar c^*} = 
-\overline{\G}^{(1;2)}_{[0]\bar c^* \bar c^*} ,
\end{align}
which is evidently satisfied. \1eq{ts.2} reads instead
\begin{align}
	\int_0^1 \dt \, t\left( 2\, \overline{\G}^{(1;2)}_{[2]\bar c^* \bar c^*} +\overline{\G}^{(1;2)}_{[1]\bar c^* \bar c^*}\right)= -\overline{\G}^{(1;2)}_{[1]\bar c^* \bar c^*},
\end{align}
which is again easily verified. A similar procedure leads to the verification of all the remaining one-loop 2-point functions.

\section{Renormalization}\label{sec.constraints}

At $z=0$ the dimension 6 operator vanishes and the theory defined by the vertex functional~\noeq{tree.level} is power-counting renormalizable as it coincides with the usual HK model. On the other hand, when $z\neq0$ power-counting renormalizability is lost, and new UV divergences appear at each loop order (starting already at one loop) proportional to increasing powers of the external momenta. The theory can be still renormalized {\it a la} Weinberg~\cite{Gomis:1995jp}: the newly appearing divergent amplitudes must be regularized by subtracting appropriate counter-terms; however, contrary to the renormalizable case, the corresponding {\it finite parts} of these amplitudes must also be fixed by appropriate renormalization conditions, reflecting the well-known fact that, being non renormalizable at $z\neq 0$, the vertex functional~\noeq{tree.level} gives rise to an effective field theory~\cite{Buchmuller:1985jz,Grzadkowski:2010es}.

At $z\neq0$ therefore a general amplitude can be decomposed as follows
\begin{align}
	\G^{(n)}_{\Phi_1 \cdots \Phi_r} = \sum_{\ell \geq 0}
    \left [ \G^{(n;\ell)}_{\Phi_1 \dots \Phi_r}
    - \sum_{k=1}^n  \frac{1}{\epsilon^k}\cntrm^{(n;\ell)}_{k;\Phi_1 \dots \Phi_r} + 
    %-\overline\G^{(n;\ell)}_{\Phi_1 \dots \Phi_r}+
    \finite^{(n;\ell)}_{\Phi_1 \dots \Phi_r}\right ],
    \label{eq.rep}
\end{align}
where: $\G^{(n;\ell)}_{\Phi_1 \cdots \Phi_r}$ is the $D$-dimensional regularized $n$-th order amplitude int the $\ell$-sector after the insertion of the counter-terms up to order $n-1$ in the loop expansion;  $\cntrm^{(n;\ell)}_{k;\Phi_1 \dots \Phi_r} $ are the residues of the poles in $1/\epsilon^k$ in the expansion of $\G^{(n;\ell)}_{\Phi_1 \dots \Phi_r}$  around $D=4$; and $\finite^{(n;\ell)}_{\Phi_1 \dots \Phi_r}$ are the finite counter-terms that inherit the degree $\ell$ from  the number of internal $X_2$-lines. Notice in particular that both $\cntrm$ and $\finite$ are Lorentz-covariant polynomials of degree $\delta_r$ (the UV degree of divergence of the corresponding amplitude) in the external momenta $p_i$; and that they obey the same differential equation~\noeq{eq.1pi.exp} of the corresponding amplitudes, namely
\begin{align}
	{\cal D}^{M^2}_z \cntrm^{(n;\ell)}_{k;\Phi_1 \dots \Phi_r} &= -
    \ell\cntrm^{(n;\ell)}_{k;\Phi_1 \dots \Phi_r};&
    {\cal D}^{M^2}_z \finite^{(n;\ell)}_{\Phi_1 \dots \Phi_r} &= -
    \ell\finite^{(n;\ell)}_{\Phi_1 \dots \Phi_r},
\end{align}
and thus possess the structure~\noeq{genstr} for their general solutions:
\begin{subequations}
\begin{align}
	\cntrm^{(n;\ell)}_{k;\Phi_1 \dots \Phi_r}(z,M^2)&=\frac1{(1+z)^\ell}\cntrm^{(n;\ell)}_{k;\Phi_1 \dots \Phi_r}(0,M^2/1+z),\\
	\finite^{(n;\ell)}_{\Phi_1 \dots \Phi_r}(z,M^2)&=\frac1{(1+z)^\ell}\finite^{(n;\ell)}_{\Phi_1 \dots \Phi_r}(0,M^2/1+z).	
\end{align}  
\end{subequations} 

Now, consider for example the case of the one-loop three-point $\sigma$ 1-PI amplitude where one has contributions from diagrams with $\ell=0,\ 1,\ 2$ and $3$. Thus we have
\begin{align}
	\G^{(1)}_{\sigma_1\sigma_2\sigma_3} = \sum_{\ell=0}^3
    \left [ \G^{(1;\ell)}_{\sigma_1\sigma_2\sigma_3}
    -\frac{1}{\epsilon}  \cntrm^{(1;\ell)}_{1;\sigma_1\sigma_2\sigma_3} + 
    %-\overline\G^{(n;\ell)}_{\Phi_1 \dots \Phi_r}+
    \finite^{(1;\ell)}_{\sigma_1\sigma_2\sigma_3}\right ],
    \label{eq.3sigma}
\end{align} 
where the UV divergent and finite parts $\cntrm$ and $\finite$ are in this case polynomials of degree 2 in the independent momenta $p_{1,2}$ ($p_3=-p_1-p_2$), or
\begin{subequations}
\begin{align}
	\cntrm^{(1;\ell)}_{1;\sigma_1\sigma_2\sigma_3}&=\ccntrm^{0(1;\ell)}_{1;\sigma_1\sigma_2\sigma_3}+\ccntrm^{1(1;\ell)}_{1;\sigma_1\sigma_2\sigma_3}(p_1^2+p_2^2+p_1{\cdot}p_2),\\
	\finite^{(1;\ell)}_{\sigma_1\sigma_2\sigma_3}&=\ffinite^{0(1;\ell)}_{\sigma_1\sigma_2\sigma_3}+\ffinite^{1(1;\ell)}_{\sigma_1\sigma_2\sigma_3}(p_1^2+p_2^2+p_1{\cdot}p_2).
\end{align}	
\end{subequations}
An explicit calculation in the Feynman gauge then yields 
\begin{subequations}
\begin{align}
    \ccntrm^{0(1;0)}_{1;\sigma_1\sigma_2\sigma_3} &= -\frac{3}{16 \pi^2 v^3}\left( m^4 - 2 m^2 M_A^2 + 12 M_A^4 \right), \\ 
    \ccntrm^{0(1;1)}_{1;\sigma_1\sigma_2\sigma_3} &= -\frac{3 M^2}{4\pi^2 v^3 (1+z)^2} \left ( m^2 + \frac{2 M^2}{1+z} \right), \\
    \ccntrm^{0(1;2)}_{1;\sigma_1\sigma_2\sigma_3}& = \frac{9}{16\pi^2 v^3 (1+z)^2}
    \left[ 
    m^4 + \frac{8 m^2 M^2}{1+z} +  \frac{12 M^4}{(1+z)^2}
    \right],\\
    \ccntrm^{0(1;3)}_{1;\sigma_1\sigma_2\sigma_3} &= -\frac{3}{4\pi^2 v^3 (1+z)^3}
    \left[ 
    m^4 + \frac{6 m^2 M^2}{1+z} +  \frac{8 M^4}{(1+z)^2}
    \right],\\
    \ccntrm^{1(1;0)}_{1;\sigma_1\sigma_2\sigma_3} &=
    % \gamma^{(1;0)}_2 = \gamma^{(1;0)}_3 = 
    0, \\
     \ccntrm^{1(1;1)}_{1;\sigma_1\sigma_2\sigma_3} &= %\gamma^{(1;1)}_2 = \gamma^{(1;1)}_3 =
    -\frac{M^2}{2 \pi^2 v^3 (1+z)^2},\\
     \ccntrm^{1(1;2)}_{1;\sigma_1\sigma_2\sigma_3} &= 
    %\gamma^{(1;2)}_2 = \gamma^{(1;2)}_3 = 
    \frac{1}{8 \pi^2 v^3 (1+z)^2}\left(m^2 +  \frac{10 M^2}{1+z}\right)
    %\left [ m^2 + \frac{10 M^2}{1+z} \right ] 
    ,\\
    \ccntrm^{1(1;3)}_{1;\sigma_1\sigma_2\sigma_3} &=
    %\gamma^{(1;3)}_2= \gamma^{(1;3)}_3= 
    -\frac{1}{8 \pi^2 v^3 (1+z)^3}\left(m^2 +  \frac{6 M^2}{1+z}\right).
\end{align}
\end{subequations}
In particular, notice that summing the different layers in $\ell$ (like one would do in a ``standard'' approach) gives for the coefficient of the quadratic term in the independent momenta
\begin{align}
    \ccntrm^{1(1)}_{1;\sigma_1\sigma_2\sigma_3}=\sum_{\ell=0}^3 \ccntrm^{1(1;\ell)}_{1;\sigma_1\sigma_2\sigma_3} & = 
    \frac{z}{8 \pi^2 v^2 (1+z)^4}\left [ 2 M^2 (1-2z) + m^2 (1 +z) \right].
    \label{gamma1}
\end{align}

Consider then first the $z=0$ case, where, without loss of generality, we can work in the Minimal Subtraction (MS) scheme (in fact, amplitudes at $z=0$ in any other subtraction scheme can be reduced to those in the MS scheme by a suitable redefinition of fields and coupling constants). As expected for a power-counting renormalizable model like the HK, \1eq{gamma1} vanishes when $z=0$; and therefore we obtain the condition on the finite parts
\begin{align}
	\left.\ffinite^{1(1)}_{1;\sigma_1\sigma_2\sigma_3}\right|_{z=0}=\sum_{\ell=0}^3 \left.\ffinite^{1(1;\ell)}_{1;\sigma_1\sigma_2\sigma_3}\right|_{z=0} & =0. 
	\label{finiteptsz=0}
\end{align}   

When $z\neq0$, \1eq{gamma1} is non-vanishing and, within an effective field theory approach, we need to impose a renormalization condition for this new term without spoiling the ST identities. This can be achieved by fixing the coefficient 
%$C^{1(1)}_1$ 
$\lambda_6^{(1)}$
of the invariant, see Eq.(\ref{g.invs})
\begin{align}
    \lambda_6^{(1)} \int \d \left( \phi^\dagger \phi - \frac{v^2}{2} \right)
    \Big ( 
    \phi^\dagger D^2 \phi
    + (D^2\phi)^\dagger \phi ) &\supset v \lambda_6^{(1)} \int \d \, \sigma^2 \square \sigma;&
    f^{1(1)}_{1;\sigma_1\sigma_2\sigma_3}\equiv 2 v \lambda_6^{(1)}.
\end{align}

However, this fixes only the overall sum over $\ell$; its decomposition in terms of $\ell$ is, in general, not uniquely determined. If, on the other hand, each $\ell$-sector could be proven to be separately Slavnov-Taylor invariant, each of these subsectors would reproduce in the limit $z \rightarrow 0$ the corresponding subsectors of the amplitude in the power-counting renormalizable theory at $z=0$. This implies that the condition~\noeq{finiteptsz=0} can be imposed at $z\neq0$
\begin{align}
	\ffinite^{1(1)}_{1;\sigma_1\sigma_2\sigma_3}=\sum_{\ell=0}^3 \ffinite^{1(1;\ell)}_{1;\sigma_1\sigma_2\sigma_3} & =0, 
	\label{finiteptsz}
\end{align}
and at this point it is immediate to prove that the condition above implies upon repeated application of the differential operator ${\cal D}^{M^2}_z$ that
\begin{align}
	\ffinite^{1(1;\ell)}_{1;\sigma_1\sigma_2\sigma_3} & =0 \quad \forall\, \ell. 
\end{align}
This result is in fact generic and not limited to the three-point $\sigma$ amplitude considered here for illustrative purposes; that is, within this formulation of the HK model at $z\neq 0$ finite parts are unambiguously set to zero by imposing the condition
\begin{align}
	&\sum_{\ell\geq0}\finite^{(n;\ell)}_{\Phi_1 \dots \Phi_r}=0; &\Longrightarrow	& &\finite^{(n;\ell)}_{\Phi_1 \dots \Phi_r}=0 \quad \forall\, \ell.
	\label{rcon}
\end{align}

In the next Section we are going to prove that fixed $\ell$ sectors are indeed separately ST-invariant. Before doing that however, let us observe that the widely studied choice of normalization conditions for the St\"uckelberg theory (the $\ell=0$ sector) in which one requires the matching of the 1-PI amplitudes of the St\"uckelberg model with those of the Higgs theory (seen as a UV completion of the former)~\cite{Dittmaier:1995cr,Dittmaier:1995ee,Herrero:1993nc} breaks the condition~\noeq{rcon}: in fact, it requires to choose the finite parts of $\finite^{(n;0)}_{\Phi_1 \dots \Phi_r}$  in such a way that the St\"uckelberg and Higgs amplitudes at $z=0$ coincide at some IR scale $\mu^2$ in the Taylor expansion up to the superficial degree of UV divergence 
$\delta_r$ of the St\"uckelberg amplitude.

In order to enforce this matching condition we need to fine-tune $\finite^{(n;0)}_{\Phi_1 \dots \Phi_r}$ in~\1eq{eq.rep} according to (all amplitudes in the r.h.s. are understood to be evaluated at $z=0$)
\begin{align}
   \left . \finite^{(n;0)}_{\Phi_1 \dots \Phi_r} \right |_{ p^2  =\mu^2} & =
   t^{\delta_r} \left ( \G^{(n)}_{\Phi_1 \dots \Phi_r} - 
   \G^{(n;0)}_{\Phi_1 \dots \Phi_r}
   + \sum_{k=1}^n \frac{1}{\epsilon^k}
   \cntrm^{(n;0)}_{k;\Phi_1 \dots \Phi_r}
   \right ),
   \label{eq.matching}
   \end{align}
where $t^{\delta_r}$ denotes the Taylor expansion up to order $\delta_r$ around a symmetric point $p^2  =\mu^2$ in the momenta $p_1, \dots, p_r$. Then, using~\1eq{eq.rep}, we obtain
\begin{align}
\left . \finite^{(n;0)}_{\Phi_1 \dots \Phi_r} \right |_{p^2  =\mu^2}& = t^{\delta_r} \left [
   \sum_{\ell \geq 1}
   \left (
   \G^{(n;\ell)}_{\Phi_1 \dots \Phi_r}
   - \sum_{k=1}^n \frac{1}{\epsilon^k} \cntrm^{(n;\ell)}_{k;\Phi_1 \dots \Phi_r} \right )
   \right ]
   + \sum_{\ell \geq 1} 
   \left . 
   \finite^{(n;\ell)}_{\Phi_1 \dots \Phi_r}
   \right |_{ \underline{p}^2  =\mu^2}, 	
\end{align}
where the last term in the above equation is a polynomial of degree $\delta_r$ in the momenta. In addition, the sum rule at $z=0$
\begin{align}
    \sum_{\ell \geq 0} \left .
    \finite^{(n;\ell)}_{\Phi_1 \dots \Phi_r} \right |_{z=0}= 0,
    \label{eq.sumrule}
\end{align}
must also hold true, since, as mentioned above, we can assume, without loss of generality, that the power-counting renormalizable HK model at $z=0$ is defined in the MS scheme. No unique choice of finite parts fulfilling simultaneously the two conditions \2eqs{eq.matching}{eq.sumrule} exists: at least two $\finite^{(n;\ell)}_{\Phi_1 \dots \Phi_r}$ must be different from zero with the remaining parts that can be arbitrarily fixed.

\section{$\ell$-sector Slavnov-Taylor identities}\label{sec.sti}

The 1-PI
amplitudes involving at least one external $X_2$ leg are uniquely fixed by the $X_2$-equation of motion Eq.(\ref{X2.eq}), so that we can concentrate on $X_2$-independent amplitudes.
At order $n$ in the loop expansion, the 1-PI amplitudes of the functional $\G(t)$ of~\1eq{Gammat} at $X_2=0$ can be gathered into a vertex functional $\widehat \Gamma(t)$ defined as 
\begin{align}
    \widehat \Gamma(t)^{(n)} \equiv
    \left .
    \G_0^{(n)}
    \right |_{X_2=0} +
    \left. t \left [ \Gamma(t) ^{(n)}- \G_0^{(n)} \right ]
    \right |_{X_2=0}
    = 
    \left .
    \G_0^{(n)} 
    \right |_{X_2=0}
    + 
    \left .
    \sum_{\ell \geq 1} t^\ell \G^{(n;\ell)}
        \right |_{X_2=0}.
    \label{mod.vf}
\end{align}

$\widehat \Gamma(t)^{(n)}$ can be recovered from the vertex functional of the complete theory $\Gamma(1)$ upon rescaling the
internal $X_2$-propagators
according to
\begin{align}
\Delta_{X_2 X_2}(t)=
\frac{i t}{(1+z)p^2 - M^2} = 
\frac{i}{\frac{1+z}{t} p^2 - \frac{M^2}{t}}\, ,
\label{X2.prop.resc}
\end{align}
i.e.
by rescaling the parameters 
$z, {M}^2$ according to
\begin{align}
    1+z \rightarrow \frac{1+z}{t} \, , \quad 
    {M}^2 \rightarrow \frac{M^2}{t} \, .
    \label{param.rescaling}
\end{align}
Since both $M^2$ and $z$ at tree-level only appear in the quadratic part, 
the choice in the above equation
entails that the parameter $t$ only enters in the bilinear term and does not affect the interaction vertices.
Moreover,
 since $X_2$ is gauge- and BRST-invariant, this choice does not violate the ST identity.
 Therefore at the regularized level the ST identity holds true for the vertex functional $\widehat \Gamma(t)$ in~\1eq{mod.vf}. 

We can obtain a local scaling equation by applying the operator ${\cal D}^{M^2}_z$ to \1eq{mod.vf}, since
\begin{align}
    {\cal D}^{M^2}_z \Gwt(t) =
    \sum_{\ell \geq 1} t^l {\cal D}^{M^2}_z  \G^{(n;l)} = -
    \sum_{\ell \geq 1} \ell t^\ell\G^{(n;\ell)} =
    - t \frac{\partial}{\partial t} \Gwt(t)
\end{align}
so that
\begin{align}
    \left (  {\cal D}^{M^2}_z + t \frac{\partial}{\partial t} \right ) \Gwt(t) = 0  \, .
\end{align}
The most general solution of this equation has the form 
\begin{align}
    \Gwt (t)= \Gwt \left ( \frac{t}{M^2}, \frac{t}{1+z} \right ) \, .
    \label{Gwt.sol}
\end{align}

Diagrams involving
only tree-level interaction vertices and rescaled internal $X_2$-lines obviously fulfill
Eq.(\ref{Gwt.sol}). 
The fact that Eq.(\ref{Gwt.sol})
holds true at the renormalized level implies that the rescaling in Eq.~(\ref{param.rescaling}) survives quantization, namely that also the counter-terms are consistent with Eq.(\ref{Gwt.sol}). 

At $t=0$ one recovers the St\"uckelberg sector of the theory; in addition, since the vertex functional $\Gwt(t)$ admits a Taylor expansion in powers of $t$, \1eq{Gwt.sol} entails that each coefficient 
$\left . \G^{(n;\ell)} \right |_{X_2=0}$
of order $t^\ell$ satisfies
the eigenvalue equation
\begin{align}
{\cal D}^{M^2}_z \left . \G^{(n;\ell)}\right |_{X_2=0} 
&= - \ell\left . \G^{(n;\ell)}\right |_{X_2=0} 
\end{align}
which is consistent with Eq.(\ref{eq.1pi.exp}).

The subtraction of UV divergences by local counter-terms with the boundary condition in \1eq{rcon} does not violate the ST identity, so that we can write at the renormalized level
\begin{align}
 {\cal S}(  \widehat \Gamma(t) ) =  0.
\end{align}
In particular, notice that since $\widehat \G(0) = \G_0$ and $\widehat \G(1) = \G(1)$, the functional $\widehat \Gamma(t)$ interpolates between the St\"uckelberg theory and the fully deformed Abelian HK model at $z \neq 0$.

We can now expand the ST identity for $\widehat \Gamma(t)$ in the number of loops and then order by order in $t$. This double expansion yields relations among 1-PI Green's functions that hold true separately. Let us then start at one loop order and use the mass eigenstate basis, that is reflected in the shift (at $\xi \neq 0$)
$b = b'+ \frac{1}{\xi} \partial A + ev \chi \, .$
We consider only amplitudes that are independent of $X_{1,2}$ since these amplitudes are recovered by the $X_{1,2}$-equations (\ref{X1.eq}) and (\ref{X2.eq}), so we can safely use $\sigma$ in place of $\sigma'$. The ST identity becomes
\begin{align}
        {\cal S}_0 (   \Gwt^{(1)} (t) ) = 
         \int \mathrm{d}^4x  \, & \left [ 
	\partial_\mu \omega \frac{\delta \Gwt^{(1)} (t)}{\delta A_\mu} 
	+ 
	\frac{\delta \G^{(0)}}{\delta \sigma^*} \frac{\delta \Gwt^{(1)}(t)}{\delta \sigma}
	+
	\frac{\delta \G^{(0)}}{\delta \sigma}  \frac{\delta \Gwt^{(1)}(t)}{\delta \sigma^*} \right.
	\nonumber \\
	& \left.
	%%%%%%%%%%%
	+ 
	\frac{\delta \G^{(0)}}{\delta \chi^*} \frac{\delta \Gwt^{(1)}(t)}{\delta \chi} +
    \frac{\delta \G^{(0)} }{\delta \chi} \frac{\delta \Gwt^{(1)}(t)}{\delta \chi^*} 
    %\nonumber \\
	%%%%%
	+ \left ( b'+ \frac{1}{\xi} \partial A + ev \chi \right ) \frac{\delta \Gwt^{(1)}(t)}{\delta \bar \omega} \right ] = 0.
    \label{sti.1loop}
\end{align}

At higher orders in the loop expansion one must take into account the effects of the bilinear antifield-dependent terms in the ST identity \noeq{sti}. Defining the bracket
\begin{align}
    (\Gwt(t),\Gwt(t)) \equiv \int \mathrm{d}^4x  \,
    \left [
    \frac{\delta \Gwt(t)}{\delta \sigma^*} 
    \frac{\delta \Gwt(t)}{\delta  \sigma} + 
    \frac{\delta \Gwt(t)}{\delta  \chi^*}
    \frac{\delta \Gwt(t)}{\delta  \chi}
    \right ],
\end{align}
at order $n>1$ in the loop expansion the ST identity yields
\begin{align}
    {\cal S}_0(\Gwt^{(n)}(t)) +
    \sum_{j=1}^{n-1}
    (\Gwt^{(j)}(t),\Gwt^{(n-j)}(t)) = 0.
\end{align}
A further expansion in powers of $t$ yields a set of independent identities valid at order $n$ one for each $\ell$ :
\begin{align}
    {\cal S}_0(\G^{(n;\ell)}) +
    \sum_{j=1}^{n-1} \sum_{i=0}^\ell
    (\G^{(j;i)},\G^{(n-j;\ell-i)}) = 0.
    \label{sti.n.k}
\end{align}
Such identities encode the conditions required to guarantee physical unitarity of the theory (i.e., the cancellation of the intermediate ghost states).

In Landau gauge major simplifications arise since the bilinear term vanishes, as a consequence of the fact that at order $n\geq 1$ amplitudes with at least one antifield external leg are zero, since there are no Feynman diagrams contributing to them. For the same reason
\begin{align}
    {\cal S}_0 (\G^{(n;\ell)}) = s (\G^{(n;\ell)}),
\end{align}
implying that there are no radiative corrections to the classical BRST symmetry. The latter is a well-known property of the Landau gauge~\cite{Blasi:1990xz,Quadri:2021syf,Grassi:1997mc}.

In particular, gauge invariance of the $X_2$-field entails that off-shell 1-PI subdiagrams with a given number of internal $X_2$-lines form separately gauge-invariant sectors. The lowest sector is the St\"uckelbeg theory (no physical $X_2$-states). Then perturbation theory based on the classical action \noeq{tree.level} will generate the higher order $\ell$-sectors (with one, two, etc. internal $X_2$-lines).

The interplay between the loop expansion and the grading in the number of internal $X_2$-lines in an arbitrary $R_\xi$-gauge requires instead to take into account the deformation of the classical BRST symmetry via the renormalization of antifield-dependent amplitudes and is encoded inthe ST identity in \1eq{sti.n.k}. It should be stressed that the grading of the ST identity with respect to the number of internal $X_2$-lines does not depend on the $z$-differential equation and the specific quadratic deformation controlled by $z$, yet it holds for any $X_2$-potential provided that $X_2$ remains gauge-invariant.

\subsection{2-point functions}
As a specific example of the ST identity just derived, let us  consider the 2-point sector. By differentiating \1eq{sti.1loop} with respect to $\omega$ and $A_\nu$, one finds
\begin{align}
    -\partial_\mu \Gwt^{(1)}_{A_\nu A_\mu}(t) + 
    \G^{(0)}_{\omega \chi^*} \Gwt^{(1)}_{A_\nu \chi}(t) + 
    \G^{(0)}_{A_\nu \chi}  \Gwt^{(1)}_{\omega \chi^*}(t) - 
    \frac{1}{\xi} \partial_\nu \Gwt^{(1)}_{\omega \bar \omega}(t) = 0.
    \label{st.aomega}
\end{align}
In a similar fashion by differentiating \1eq{sti.1loop} with respect to $\omega,\chi$ one gets
\begin{align}
    -\partial_\mu \Gwt^{(1)}_{\chi A_\mu}(t) + 
    \G^{(0)}_{\omega \chi^*} \Gwt^{(1)}_{\chi \chi}(t) + \G^{(0)}_{\omega \sigma^* \chi} \Gwt^{(1)}_{\sigma}(t) +
    \G^{(0)}_{\chi \chi}  \Gwt^{(1)}_{\omega \chi^*}(t) + 
    ev  \Gwt^{(1)}_{\omega \bar \omega}(t) = 0 \, .
    \label{st.chiomega}
\end{align}
In Landau gauge instead no shift of the $b$-field is required and the ST identities for the two point functions are simpler (notice that  the one-loop amplitudes with external antifield legs are zero in this gauge):
\begin{align}
    &  \left .  -\partial_\mu \Gwt^{(1)}_{A_\nu A_\mu}(t) \right |_{\xi=0}  + 
    \G^{(0)}_{\omega \chi^*}  \left .  \Gwt^{(1)}_{A_\nu \chi}(t)  \right |_{\xi=0}  = 0 \, , \nonumber \\
    & 
        \left .  -\partial_\mu \Gwt^{(1)}_{\chi A_\mu}(t)  \right |_{\xi=0}  + 
    \G^{(0)}_{\omega \chi^*}  \left .  \Gwt^{(1)}_{\chi \chi}(t) \right |_{\xi=0}  + \G^{(0)}_{\omega \sigma^* \chi}  \left .  \Gwt^{(1)}_{\sigma}(t)  \right |_{\xi=0} = 0 \, .
    \label{st.2pt.l}
\end{align}

We now project \3eqs{st.aomega}{st.chiomega}{st.2pt.l} at zero and first order in $t$ (there are no further contributions since the amplitudes involve at most one internal $X_2$-line). In the Feynman gauge we then obtain ($\ell=0,1$):
\begin{subequations}
\begin{align}
& 
    \left . -\partial_\mu \G^{(1;\ell)}_{A_\nu A_\mu} \right |_{\xi=1} + 
    \G^{(0)}_{\omega \chi^*} 
     \left .
    \G^{(1;\ell)}_{A_\nu \chi}
     \right |_{\xi=1} + 
    \G^{(0)}_{A_\nu \chi}  
    \left . 
    \G^{(1;\ell)}_{\omega \chi^*}
     \right |_{\xi=1} - 
   \left .  \partial_\nu
   \G^{(1;\ell)}_{\omega \bar \omega} \right |_{\xi=1} = 0 ,  \\
& 
 -\partial_\mu 
 \left .
 \G^{(1;\ell)}_{\chi A_\mu}
  \right |_{\xi=1} + 
    \G^{(0)}_{\omega \chi^*} 
    \left . \G^{(1;\ell)}_{\chi \chi} \right |_{\xi=1} + 
    \G^{(0)}_{\chi \chi} \left . \G^{(1;\ell)}_{\omega \chi^*} \right |_{\xi=1}  + 
     \left .  \G^{(0)}_{\omega \sigma^* \chi} \G^{(1;\ell)}_{\sigma} \right |_{\xi=1} +
    ev \left . \G^{(1;\ell)}_{\omega \bar \omega}
     \right |_{\xi=1} = 0, 
\end{align}	
\end{subequations}
whereas in the Landau gauge we get
\begin{subequations}
\begin{align}
&     -\partial_\mu \left . \G^{(1;\ell)}_{A_\nu A_\mu} \right |_{\xi=0} + 
    \G^{(0)}_{\omega \chi^*} 
     \left .
     \G^{(1;\ell)}_{A_\nu \chi}
     \right |_{\xi=0}   = 0, 
     \\
&    -\partial_\mu \left . \G^{(1;\ell)}_{\chi A_\mu} \right |_{\xi=0} + 
    \G^{(0)}_{\omega \chi^*} 
    \left . 
    \G^{(1;\ell)}_{\chi \chi} \right |_{\xi=0} +
    \left .  \G^{(0)}_{\omega \sigma^* \chi}  \G^{(1;\ell)}_{\sigma} \right |_{\xi=0} = 0.
\end{align}
\label{st.2pts.k}
\end{subequations}
At variance with the standard $\phi$-formalism, the identities above hold true separately for each sector with a given number of internal $X_2$-lines. They can be explicitly checked using the following results:
\begin{align}
	   & \overline{\G}^{(1;0)}_{A_\mu A_\nu} =
    - \frac{e^2 M_A^2}{8 \pi^2}\frac{g^{\mu\nu}}{\epsilon} \delta_{\xi;1}; 
    \qquad
    \overline{\G}^{(1;1)}_{A_\mu A_\nu} =
     \frac{e^2 }{24 \pi^2 (1+z)} \left [ 
    -( 9 M_A^2 + p^2 ) 
    g^{\mu\nu} + p^\mu p^\nu
    \right ] \frac{1}{ \epsilon},\nonumber \\
     & \overline{\G}^{(1;0)}_{\chi\chi} = \frac{ M_A^2}{16 \pi^2 v^2}
     \left [ 
     6 M_A^2 + 
      \delta_{\xi;1}
     (m^2 - 2 p^2)
     \right ]
     \frac{1}{\epsilon},
     \nonumber \\
     & 
     \overline{\G}^{(1;1)}_{\chi\chi} = \frac{1}{
     16 \pi^2 v^2 (1+z)^3}
     \Big \{ 
     M^2 [ (1+z) m^2 +2 M^2 ]
     - 2 ( 3 -2 \delta_{\xi;1} )
     M_A^2 (1+z)^2  p^2
     \Big \}
     \frac{1}{\epsilon},
     \nonumber \\
     & \overline{\G}^{(1;0)}_{\chi A_\mu(p)} = i
     \frac{e M_A^2}{8 \pi^2 v} \frac{\delta_{\xi;1}}{\epsilon} p^\mu;
     \qquad
     \overline{\G}^{(1;1)}_{\chi A_\mu(p)} = i
     \frac{e M_A^2 ( 3 \delta_{\xi;0} +  2 \delta_{\xi;1}) }{8 \pi^2 (1+z) v}\frac{1}{\epsilon} p^\mu,
     \nonumber \\
     & \overline{\G}^{(1;0)}_{\omega \bar \omega} = 0 \, ; \quad
     \overline{\G}^{(1;1)}_{\omega \bar \omega  } = - \frac{e^2 M_A^2}{8 \pi^2 (1+z)}\frac{\delta_{\xi;1}}{\epsilon};
     \quad
     \overline{\G}^{(1;0)}_{\omega \chi^* } = 0 \, ; \quad
     \overline{\G}^{(1;1)}_{\omega \chi^*  } = -\frac{e^2 M_A}{8 \pi^2 (1+z)}\frac{\delta_{\xi;1}}{\epsilon},
\end{align}
where $\delta_{\xi;1}$ ($\delta_{\xi;0}$) is 1 in the Feynman (Landau) gauge and zero otherwise.

\section{Other renormalization schemes}\label{sec:otherrens}

Separate invariance under the ST identities for each $\ell$-sector has some important consequences. For example, when chiral fermions are introduced, as in the Standard Model, the presence of the $\gamma_5$-matrix entails that the ST identities are broken by the intermediate regularization and that finite counter-terms must be added to recover the validity of the ST identities themselves. In order to be consistent with the $z$-differential equation, such finite counter-terms must also possess a grading in the number of internal $X_2$-lines; and the fact that for each $\ell$-sector a separate ST identity exists implies that the breaking of the $\ell$-th ST identity can be compensated by finite counter-terms also belonging to the same $\ell$-sector.

Suppose now that at $z=0$ the ST identities the $\ell$-sector are fulfilled, namely
\begin{align}
    {\cal S}_0(\left . \G^{(n;\ell)} \right |_{z=0}) +
    \sum_{j=1}^{n-1} \sum_{i=0}^\ell
    (\left . \G^{(j;i)} \right |_{z=0},\left . \G^{(n-j;\ell-i)} \right |_{z=0}) = 0.
    \label{sti.l.exp}
\end{align}
Then, since in both terms in Eq.(\ref{sti.l.exp}) a common overall factor 
$1/(1+z)^\ell$ can be factorized and 
\begin{align}
    \left [ {\cal S}_0,\frac{\partial}{\partial {M^2}} \right ] =
    \left [ {\cal S}_0,\frac{\partial}{\partial z} \right ] = 0.
\end{align}
a solution to the ST identities at $z \neq 0$ is provided by
\begin{align}
   \G^{(n;\ell)}(z,M^2) = \frac{1}{(1+z)^\ell}\G^{(n;\ell)}(0,M^2/1+z).
   \label{solution}
\end{align}
Similarly, all the other functional identities of the theory, {\it e.g.}, the one reported in Eqs.(\ref{sti.c})-(\ref{X2.eq}), (\ref{b.eq}) and (\ref{antigh.eq}), can be expanded in the different $\ell$-sectors and each projected equation is satisfied by $\G^{(n;\ell)}$ in~\1eq{solution} if $\left . \G^{(n;\ell)} \right |_{z=0}$ is a solution of the same equation at $z=0$.

As \1eq{solution} holds at the renormalized level, it can be taken as the definition of the 1-PI amplitudes of the theory at $z\neq 0$. Within such a formulation, one can then extend the construction presented to an arbitrary renormalization scheme of the power-counting renormalizable theory at $z=0$, {\it e.g.}, the on-mass shell renormalization scheme. In fact finite renormalizations compatible with the symmetries of the theory are encoded into the relevant amplitudes contributing to $\G^{(n;\ell)}(0,M^2)$ and  are lifted at $z\neq 0$ by Eq.(\ref{solution}), as such normalization conditions inherit (as a consequence of the validity of the ST identities in each $\ell$-sector) a natural $\ell$ grading.

Consider as an example the on-mass shell normalization condition for the vector meson. Defining
\begin{eqnarray}
\G_{A^\mu A^\nu} = g_{\mu\nu} (p^2 - M_A^2)
+ \left ( g_{\mu\nu} - \frac{p_\mu p_\nu}{p^2} \right ) \Sigma_T(p^2) + \frac{p^\mu p^\nu}{p^2}
\Sigma_L(p^2)
\end{eqnarray}
one requires that the position of the pole of the physical components of the vector meson does not shift with respect to the one at tree level and that the residue of the propagator on the pole is one:
\begin{align}
    {\mbox{Re}}~ \Sigma_T(M_A^2) &= 0;&
        \left . 
    {\mbox{Re}}~\frac{\partial \Sigma_T(p^2)}{\partial p^2} \right |_{p^2=M_A^2} &= 0.
    \label{on.mass.nc}
\end{align}
These conditions can be matched by finite renormalizations involving the invariants shown in~\1eq{g.invs}: 
\begin{align}
    \lambda_4 \int \d (D^\mu \phi)^\dagger D_\mu \phi &\supset \frac{\lambda_4 v}{2} \int \d A_\mu^2;&    \frac{\lambda_8}{2} \int \d 
    F_{\mu\nu}^2 &\supset \lambda_8 \int \d
    A_\mu (\square g^{\mu\nu} - \partial^\mu \partial^\nu ) A_\nu. 
\end{align}
Then, projecting into the different $\ell$-sectors, with $\ell=0,1$, which can be done due to the $\ell$-sector ST invariance, one obtains
\begin{align}
    \mbox{Re} ~\Sigma_T^{(1;\ell)}(M_A^2) + v \lambda_4^{(1;\ell)} &= 0;&
    \mbox{Re} ~ \left . 
    \frac{\partial \Sigma_T^{(1;\ell)}}{\partial p^2}
    \right |_{p^2=M_A^2} - 2 M_A^2 \lambda_8^{(1;\ell)} &= 0. \, .
\end{align}
As can be seen from the above equation,
on mass shell renormalization conditions
respect the layers in $\ell$ of the
ST identities and consequently the
$z$-differential equation.
Once the appropriate normalization conditions
are enforced at order $n$ in the loop expansion at $z=0$, Eq.(\ref{solution})
fixes the 1-PI amplitudes of the theory
at $z\neq 0$ in a unique way.

\section{Conclusions}\label{sec:conclusions}

In the present paper we have shown that a particular dimension 6 derivative operator can be introduced in the Abelian Higgs-Kibble model in such a way that a novel differential equation uniquely fixes the dependence of the 1-PI amplitudes on the deformation parameter~$z$ and governs the subtraction of UV divergences. Power-counting renormalizabilty is lost, yet the model depends on the same number of physical parameters of the renormalizable theory at $z=0$ plus the $z$-parameter itself.

A crucial property of the theory is that the ST identity separately holds for each sector of the vertex functional with a fixed number $\ell$ of internal $X_2$-lines. This is at variance with the standard formalism, in which the dependence on $z$ affects both the bilinear and the interaction terms in the tree-level classical action (and hence no $z$-differential equation exists) and the ST identity cannot be filtered according to the number of internal physical Higgs propagators (as the physical Higgs field $\sigma$ is not gauge-invariant).

Interestingly enough, this unique deformation, controlled by the parameter $z$ and allowed by the symmetries of the theory, only affects the potential of the Higgs field. As such, it provides a candidate of a deformation of the usual quartic Higgs potential that might be relevant for the study of the electroweak spontaneous symmetry breaking. The extension to the SU(2) $\times$ U(1) gauge group and the phenomenological implications of the $z$-deformation are currently under investigation, and we hope to report soon on our findings.

\appendix

\section{The classical vertex functional in the $X$-formalism}

The vertex functional of the Higgs-Kibble model in the $X_{1,2}$ formulation reads 
\begin{align}
	\G^{(0)}  = %S + S_\mathrm{gf}  + S_\mathrm{ext} \nonumber \\
	   \int \!\mathrm{d}^4x \, &\Bigg [ -\frac{1}{4} F^{\mu\nu} F_{\mu\nu} + (D^\mu \phi)^\dagger (D_\mu \phi) - \frac{M^2-m^2}{2} X_2^2 - \frac{m^2}{2v^2} \left ( \phi^\dagger \phi - \frac{v^2}{2} \right )^2 \nonumber \\
%	&  + i \bar \psi \slashed{\cal D} \psi 
%	+ \frac{G}{\sqrt{2}} \bar{\psi} (1-\gamma_5) \psi \phi +\frac{G}{\sqrt{2}} \bar{\psi} (1 + \gamma_5) \psi \phi^\dagger
%	\nonumber \\
	&  +\frac{z}{2} \partial^\mu X_2 \partial_\mu X_2 - \bar c (\square + m^2) c + \frac{1}{v} (X_1 + X_2) (\square + m^2) \left ( \phi^\dagger \phi - \frac{v^2}{2} - v X_2 \right ) \nonumber \\
	& + \frac{\xi b^2}{2} -  b \left ( \partial A + \xi e v \chi \right ) + \bar{\omega}\left ( \square \omega + \xi e^2 v (\sigma + v) \omega\right ) \nonumber \\
%	& -  i \frac{e}{2} \psi^* \gamma_5 \psi \omega
%	+ i \frac{e}{2} \bar \psi^* \omega \bar \psi \gamma_5 %
%	\nonumber \\
	&  + \bar c^* \left ( \phi^\dagger \phi - \frac{v^2}{2} - v X_2 \right ) + \sigma^* (-e \omega \chi) + \chi^* e \omega (\sigma + v) \Bigg ].
	\label{tree.level}
\end{align}
In the above equation $D_\mu$ is the covariant derivative
\begin{align}
	D_\mu = \partial_\mu - i e A_\mu .
\end{align} 
$A_\mu$ is the Abelian gauge connection, $F_{\mu\nu} = \partial_\mu A_\nu - \partial_\nu A_\mu$ is the field strength. $\phi$ is the complex Higgs field,
\begin{align}
	\phi = \frac{1}{\sqrt{2}} ( \sigma + v + i \chi),
\end{align}
with $\chi$ is the pseudo-Goldstone field; finally, while $v$ is the vacuum expectation value.

The model in~\1eq{tree.level} contains an enlarged scalar sector with respect to the usual $\phi$-formalism controlled by the fields $X_{1,2}$. $X_1$ is a Lagrange multiplier enforcing on-shell the condition\footnote{Going on-shell with $X_1$ yields a Klein-Gordon equation
$$ (\square + m^2) \left ( \phi^\dagger \phi - \frac{v^2}{2} - v X_2 \right ) = 0, $$
so that the most general solution is $X_2 = \frac{1}{v} \left ( \phi^\dagger \phi - \frac{v^2}{2} \right ) + \eta$, $\eta$ being a scalar field of mass $m$.
However in
perturbation theory the correlators of the mode $\eta$ with any gauge-invariant operators vanish~\cite{Binosi:2019olm}, so that
one can safely set $\eta= 0$.
}
\begin{align} 
X_2 \sim \frac{1}{v} \left ( \phi^\dagger \phi - \frac{v^2}{2} \right ),
\label{os.constraint}
\end{align}
namely $X_2$ can be thought of as a field coordinate parameterizing  the scalar gauge-invariant combination $\phi^\dagger \phi - \frac{v^2}{2}$. Notice that at the linearized level $X_2 \sim \sigma$.

By eliminating in \1eq{tree.level} both $X_1$ and $X_2$ via their equations of motion one recovers the usual vertex functional of the Abelian Higgs-Kibble model with thedimension 6 derivative operator 
\begin{align}
    \frac{z}{2} \partial^\mu X_2 \partial_\mu X_2 \sim
    \frac{1}{2 v^2} \partial^\mu  \left ( \phi^\dagger \phi - \frac{v^2}{2} \right ) \partial_\mu  \left ( \phi^\dagger \phi - \frac{v^2}{2} \right ) 
\end{align}
in the usual $\phi$-formalism. In particular, the two mass parameters $m$ and $M$ in the first line of~\1eq{tree.level} are chosen in such a way that by going on-shell with the Lagrange multiplier $X_1$ one recovers the usual quartic Higgs potential
$ - \frac{M^2}{2v^2} 
 \left ( \phi^\dagger \phi - \frac{v^2}{2} \right )^2 \, .$
The only physical parameter is thus $M$. In fact, it can be checked that the correlators of physical observables after going on-shell with the fields $X_{1,2}$ do not depend on $m$~\cite{Binosi:2017ubk}.

The second line of \1eq{tree.level} contains the deformation of the $X_2$-kinetic term controlled by the parameter $z$. When such term is switched off ($z=0$), we recover the power-counting renormalizable Higgs-Kibble model; on the other hand, at $z \neq 0$, the theory becomes non-renormalizable, and is defined by solving the $z$-differential equation~\1eq{z.eq}) with appropriate boundary conditions as discussed in Sect.~\ref{sec.constraints}.

The extension of the scalar sector via the fields $X_{1,2}$ does not introduce additional physical degrees of freedom. This can be seen at tree-level by inspecting the propagators in the mass eigenstate basis, see \1eq{app.prop.1}. We notice that $\Delta_{X_1 X_1}$ and $\Delta_{\sigma'\sigma'}$ differ by a sign and they in fact cancel out in the intermediate states; this is a consequence of the constraint U(1) BRST symmetry 
\begin{align}
    \s X_1 = v c \, , \quad \s c = 0 \, , \quad
    \s \bar c = \frac{1}{v}
    \left ( \phi^\dagger \phi - \frac{v^ 2}{2} - v X_2 \right ) \, ,
    \label{brst.constr}
\end{align}
all other fields and external sources being invariant under $\s$ and $c, \bar c$ being the constraint U(1) ghost and antighost fields.

We notice that the whole dependence of the classical action~\1eq{tree.level}) on $m^2$ is BRST-exact, since
\begin{align}
&
\int \d \Big \{ 
    \frac{m^2}{2} X_2^2 - \frac{m^2}{2v^2} \left ( \phi^\dagger \phi - \frac{v^2}{2} \right )^2 + \frac{1}{v} (X_1 + X_2) (\square + m^2) \left ( \phi^\dagger \phi - \frac{v^2}{2} - v X_2 \right ) - \bar c (\square + m^2) c \Big \} \nonumber \\
& \qquad \qquad =\int \d  
    \s \left [ - \frac{m^2}{2v^2}  \bar c
    \left ( \phi^\dagger \phi - \frac{v^2}{2} 
   + v X_2
    \right )  + \frac{1}{v}
    (X_1 + X_2)
    (\square + m^2) \bar c
    \right ] \, .
\end{align}
Therefore, physical observables cannot depend on this parameter~\cite{Piguet:1995er}, in agreement with the explicit computations of~\cite{Binosi:2017ubk}.

In order to see that the physical states of the theory are unchanged with respect to the $\phi$-formalism, we make use of the BRST quantization approach~\cite{Becchi:1974xu,Becchi:1974md,Becchi:1975nq}. We remark that the BRST symmetry in \1eq{brst.constr} holds true together with  the usual BRST symmetry $s$ associated with the Abelian gauge group
\begin{align}
    s A_\mu = \partial_\mu \omega ;  \quad s\phi = i e \omega \phi  ;  \quad s \sigma = - e \omega \chi ;   \quad s \chi = e \omega ( \sigma + v) ;  
    \quad s \bar \omega = b ;  \quad s b = 0,
    \label{brst.gauge}
\end{align}
with all other fields and external sources being $s$-invariant. $\omega,\bar \omega$ are the ghost and antighost fields associated with the gauge symmetry; $b$ is the Nakanishi-Lautrup field.

Both BRST differentials $s$ and $\s$ are nilpotent and they anticommute, so that  $s' = s + \s$ is nilpotent too. Then the physical Hilbert space can be defined according to the BRST quantization prescription~\cite{Becchi:1974xu,Becchi:1974md,Becchi:1975nq}
as 
\begin{align}
    {\cal H}_{phys} = \frac{\mbox{Ker}~s'_0}{\mbox{Im}~s'_0},\end{align}
where $s'_0$ is the linearized full BRST differential acting on the fields as follows:
\begin{gather}
 s'_0 A_\mu = \partial_\mu \omega ;  \quad
  s'_0 \chi = e v \omega ;  \quad s'_0 \sigma = 0 ; 
  \quad
   s'_0 \bar \omega = b ;  \quad s'_0 b = 0 , 
  \nonumber \\
 s'_0 X_1 = v c ;  \quad s'_0 c = 0 ;  \quad s'_0 \bar c = \sigma - X_2 ;  \quad s'_0 X_2 = 0.
\end{gather}
Fron above equation one can easily see that the transverse polarizations of the massive gauge field belong to ${\cal H}_{phys}$ as well as the physical scalar $\sigma$. The latter can be equivalently parameterized at the linearized level by $\sigma$ or $X_2$, since their difference is $s'_0$-exact (being the image of $\bar c$) and thus $X_2$ and $\sigma$ belong to the same equivalence class in ${\cal H}_{phys}$. $\chi$ is unphysical since it does not belong to the kernel of $s'_0$. Moreover the ghost $\omega$ is unphysical being (modulo a constant factor) the $s'_0$-image of $\chi$; thus $(\chi, ev \omega)$ form a so-called BRST doublet and consequently they do not contribute to the cohomology of $s'_0$ that determines ${\cal H}_{phys}$ (for a review see e.g.~\cite{Barnich:2000zw}). Similarly, the pairs $(\bar \omega , b)$, $(X_1, c)$, $(\bar c, \sigma - X_2)$ form BRST doublets and thus drop out of ${\cal H}_{phys}$. We therefore conclude that the physical field content of the theory is still given by the transverse modes of the massive gauge fields and one physical massive Higgs scalar.

The last line of \1eq{tree.level} contains the antifield-dependent terms, {\it i.e.}, the terms coupling the antifields (external sources of the BRST transformation) and the non-linear BRST variations of the fields that, being non-linear, require an independent renormalization with respect to the fields themselves~\cite{Gomis:1994he,Piguet:1995er}, controlled by the anti-fields Green's functions.

\section{Propagators}

The diagonalization of the quadratic part of the classical action in the sector spanned by $\sigma,X_1,X_2$ is achieved via the field redefinition $\sigma=\sigma'+X_1+X_2$. $\sigma', X_1, X_2$ are the mass eigenstates at tree-level. Their propagators read
\begin{align}
    &\Delta_{\sigma'\sigma'} = \frac{i}{p^2-m^2}; &
    &\Delta_{X_1 X_1} = -\frac{i}{p^2-m^2};&
    &\Delta_{X_2X_2} = \frac{i}{(1+z)p^2-M^2}.
\label{app.prop.1}
\end{align}

Diagonalization in the gauge sector is obtained by redefining the Nakanishi-Lautrup multiplier field 
\begin{align}
    b'= b - \frac{1}{\xi} \partial A - ev \chi .
\end{align}
Then, the $A_\mu$-propagator is
\begin{align}
    \Delta_{\mu\nu} &= -i \left ( \frac{1}{p^2-M_A^2} T_{\mu\nu} + \frac{1}{\frac{1}{\xi} p^2-M_A^2}L_{\mu\nu}  \right );& \qquad M_A &= ev,
\end{align}
with
\begin{align}
	T_{\mu\nu} &= g_{\mu \nu} - \frac{p_\mu p_\nu}{p^2};& L_{\mu\nu} &= \frac{p_\mu p_\nu}{p^2}, 
\end{align}
whereas the Nakanishi-Lautrup, pseudo-Goldstone and ghost propagators are
\begin{align}
    \Delta_{b'b'} &= \frac{i}{\xi};& \Delta_{\chi\chi} &= \frac{i}{p^2-\xi M_A};&
    \Delta_{\bar \omega \omega} &= \frac{i}{p^2- \xi M^2_A}.
\end{align}

The Feynman gauge corresponds to $\xi =1$, whereas the Landau gauge is $\xi= 0$. Finally, the ghost associated to the constraint BRST symmetry is free with a propagator
\begin{align}
    \Delta_{\bar c c} = \frac{-i}{p^2-m^2}.
\end{align}

\section{Functional identities}

In this Appendix we collect for the sake of reference the
functional identities controlling the theory:
\begin{itemize}
    \item The ST identity for the constraint BRST symmetry is
\begin{align}
    {\cal S}_{\scriptscriptstyle{C}}(\G) \equiv \int \!\mathrm{d}^4 x \, \left [ v c \frac{\delta \G}{\delta X_1} 
 + \frac{\delta \G}{\delta \bar c^*}\frac{\delta \G}{\delta \bar c} \right ] = 
 \int \!\mathrm{d}^4 x \, \left [ v c \frac{\delta \G}{\delta X_1} 
 -(\square + m^2) c \frac{\delta \G}{\delta \bar c^*} \right ] = 0,
 \label{sti.c} 
\end{align}
where in the latter equality we have used the fact that both the ghost $c$ and the antighost $\bar c$ are free:
\begin{align}
    \frac{\delta \G}{\delta \bar c} &= -(\square + m^2) c;&     \frac{\delta \G}{\delta c} &= (\square + m^2) \bar c. 
     \label{eom.c}
\end{align}
\item The $X_1$-equation of motion, that follows from Eq.(\ref{sti.c}) by using the fact that the ghost $c$ is free:
\begin{align}
    \frac{\delta \G}{\delta X_1}=
    \frac{1}{v} (\square + m^2)
    \frac{\delta \G}{\delta \bar c^*}.    
    \label{X1.eq}
\end{align}
\item The $X_2$-equation of motion:
\begin{eqnarray}
	\frac{\delta \G}{\delta X_2} =  \frac{1}{v} (\square + m^2) \frac{\delta \G}{\delta \bar c^*} - (\square + m^2)X_1 - ( (1+z) \square + M^2) X_2 - v \bar c^*.
	\label{X2.eq}
\end{eqnarray}
Notice that the $z$-term only affects the linear contribution in the right-hand side of the above equation and thus no new external source is needed to control its renormalization.
Notice that further bilinear terms $\sim X_2 \square^n X_2, n\geq 2$ could be added to the classical action while still modifying only the linear part in $X_2$ of the above equation. However, such higher-derivative contributions induce in the spectrum modes with negative metrics and lead to mathematical inconsistencies~\cite{Aglietti:2016pwz}; thus, in this respect, the $z$-deformation of the classical HK action studied here is unique.
\item The ST identity
associated to the gauge group
BRST symmetry 
\begin{align}
	& {\cal S}(\G)  = \int \mathrm{d}^4x \, \left [ 
	\partial_\mu \omega \frac{\delta \G}{\delta A_\mu} + \frac{\delta \G}{\delta \sigma^*} \frac{\delta \G}{\delta \sigma}  + \frac{\delta \G}{\delta \chi^*} \frac{\delta \G}{\delta \chi} 
	+ b \frac{\delta \G}{\delta \bar \omega} \right ] = 0
	\label{sti} 
\end{align}
\item The $b$-equation:
\begin{eqnarray}
	\frac{\delta \G}{\delta b} =\xi b - \partial A -  \xi e v \chi.
	\label{b.eq}
\end{eqnarray}
\item The antighost equation:
\begin{eqnarray}
	\frac{\delta \G}{\delta \bar \omega} = \square \omega + \xi ev \frac{\delta \G}{\delta \chi^*}.
	\label{antigh.eq}
\end{eqnarray}   
\end{itemize}

\subsection{Descendant and Ancestor Amplitudes}

Eqs.(\ref{X1.eq}) and (\ref{X2.eq}) fix the amplitudes involving $X_1$ and $X_2$ external legs in terms of $X_{1,2}$-independent amplitudes. Hence a hierarchy arises among 1-PI Green's functions: we call amplitudes with at least one $X_1$ or $X_2$ external legs descendant amplitudes, while the amplitudes without $X_{1,2}$ external legs  are called ancestor amplitudes.

As an example, we derive the two-point $X_{1,2}$-amplitudes in terms of the 1-PI amplitudes with external $\bar c^*$-legs. We start by differentiating~\1eq{X1.eq} at order $n \geq 1 $ in the loop expansion with respect to $X_{1,2}$ we get
\begin{align}
    \G^{(n)}_{X_1 X_1} &= \frac{1}{v} (\square + m^2) \G^{(n)}_{X_1 \bar c^*};&
    \G^{(n)}_{X_2 X_1} &= \frac{1}{v} (\square + m^2) \G^{(n)}_{X_2 \bar c^*}. 
    \label{eq.x1x1}
\end{align}
Next, we differentiate~\1eq{X2.eq} at order $n \geq 1$  obtaining%
\begin{align}
     \G^{(n)}_{X_2 X_2} = \frac{1}{v} (\square + m^2) \G^{(n)}_{X_2 \bar c^*}.
    \label{eq.x2x2}
\end{align}
Finally, differentiation of \2eqs{X1.eq}{X2.eq} with respect to
$\bar c^*$ yields
\begin{align}
    \G^{(n)}_{X_1 \bar c^*} = 
    \G^{(n)}_{X_2 \bar c^*} = 
    \frac{1}{v}
    (\square + m^2) 
    \G^{(n)}_{\bar c^* \bar c^*}. 
    \label{eq.x12bc}
\end{align}
Substituting \1eq{eq.x12bc}) into~\1eq{eq.x1x1} we finally obtain
\begin{align}
    \G^{(n)}_{X_1 X_1} =
    \G^{(n)}_{X_1 X_2} =
    \G^{(n)}_{X_2 X_2}= \frac{1}{v^2} (\square + m^2)^2 \G^{(n)}_{\bar c^* \bar c^*}..
\end{align}
As anticipated, the two-point functions of the fields $X_{1,2}$ are completely fixed by the ancestor amplitude $\G^{(n)}_{\bar c^* \bar c^*}$. We can therefore limit the analysis to the $X_{1,2}$-independent 1-PI amplitudes, the latter being recovered algebraically by functional differentiation of  \1eq{X1.eq}{X2.eq}.

Next, due to the fact that the constraint ghosts 
$\bar c,c$ are free, at order $n\geq 1$ in the loop expansion the vertex functional $\G$ is $c,\bar c$-independent. We can then take a derivative w.r.t. $c$ of \1eq{sti.c} and then substitute~\1eq{eom.c} to recover the $X_{1}$-equation of motion in~\1eq{X1.eq}; thus we see that the constraint ST identity~\1eq{sti.c} is equivalent to the $X_1$-equation.

Finally, the antighost equation~\noeq{antigh.eq}) at order $n\geq 1$
\begin{eqnarray}
	\frac{\delta \G^{(n)}}{\delta \bar \omega} = \xi ev \frac{\delta \G^{(n)}}{\delta \chi^*}, 
	\label{antigh.eq.n}
\end{eqnarray}  
entails that
the 1-PI vertex functional only depends on the antighost $\bar \omega$ only via the combination $\widehat{\chi}^* = \chi^* + \xi e v \bar \omega$; and the $b$-equation~\noeq{b.eq} implies that at order $n\geq 1$ there is no dependence on the Nakanishi-Lautrup field
\begin{align}
    \frac{\delta \G^{(n)}}{\delta b} = 0.
\end{align}

\section{List of invariants}\label{app.invs}

We list here the set of invariants needed to control  the UV divergences of  operators up to dimension $6$ at one loop order.
The list is taken from Ref.~\cite{Binosi:2019nwz} upon setting to zero the source $T_1$ coupled to the derivative interaction $\sim T_1
(D^\mu \phi)^\dagger D_\mu \phi$ that is not present here.
In order to match with the results of Ref.~\cite{Binosi:2019nwz}
one must also set the coupling constant $g$ of the derivative interaction $\phi^\dagger \phi (D^\mu \phi)^\dagger D_\mu \phi$
equal to zero.
We keep the numbering used in
Ref.~\cite{Binosi:2019nwz} and 
use a bar to denote the UV-divergent part of the coefficients.

In the cohomologically trivial sector two invariants must be considered:
\begin{align}
\cfct{0}  {\cal S}_0 \! \int \d [\sigma^*(\sigma + v) + \chi^* \chi];&
    &\cfct{1}\, {\cal S}_0\! \int \d \, (\sigma^* \sigma  + \chi^* \chi).
\end{align}
%
%$In Feynman gauge one can safely set $\cfct{0}=0$~\cite{Binosi:2019nwz}.

The invariants involving the external source $\bar c^*$ are
\begin{align}
     &\cfps{1} \int \d  \bar c^*  ;
     \qquad 
     \cfps{3} \int \d \frac{1}{2}  (\bar c^*)^2  ; \qquad
     \cfps{9} \int \d \frac{1}{3!}  (\bar c^*)^3  ; 
     \label{ESinv}
\end{align}
and
\begin{align}
    & \cfxt{1} \int \d  \bar c^* \left ( \phi^\dagger \phi - \frac{v^2}{2} \right );&  
    & \cfxt{3} \int \d  \bar c^* (D^\mu \phi)^\dagger D_\mu \phi;& \nonumber \\
    & \cfxt{5} \int \d  \bar c^* \left [ (D^2 \phi)^\dagger  \phi + \mathrm{h.c.} \right ];&  
    & \cfxt{7} \int \d  \bar c^* \left ( \phi^\dagger \phi - \frac{v^2}{2} \right )^2 ;&  
    \nonumber \\
    & \cfxt{9} \int \d  \bar c^*
    F_{\mu\nu}^2; &
    & \cfxt{13} \int \d 
    (\bar c^*)^2 \left ( \phi^\dagger \phi - \frac{v^2}{2} \right ).&
    \label{mix.invs}
 \end{align}
The invariants only involving the fields of the theory are
\begin{align}
&  \cfgi{1} \int \d \left (
    \phi^\dagger \phi - \frac{v^2}{2}
    \right );&
& \cfgi{2} \int \d \left (
    \phi^\dagger \phi - \frac{v^2}{2}
    \right )^2; \nonumber \\
& \cfgi{3} \int \d \left (
    \phi^\dagger \phi - \frac{v^2}{2}
    \right )^3; &
& \cfgi{4} \int \d 
   (D^\mu \phi)^\dagger D_\mu \phi; \nonumber \\    
& \cfgi{5} \int \d \phi^\dagger
  [ (D^2)^2 + 
  D^\mu D^\nu D_\mu D_\nu + 
  D^\mu D^2 D_\mu ] \phi;& 
& \cfgi{6} \int \d \left ( \phi^\dagger \phi
  - \frac{v^2}{2} \right ) 
  \left ( \phi^\dagger D^2 \phi + 
  (D^2\phi)^\dagger \phi \right );
  \nonumber \\
& \cfgi{7} \int \d 
\left ( \phi^\dagger \phi
  - \frac{v^2}{2} \right ) (D^\mu \phi)^\dagger D_\mu \phi; &
  & \frac{\cfgi{8}}2 \int \d F_{\mu\nu}^2  ; \nonumber \\
 & \cfgi{9}  \int \d \partial^\mu F_{\mu\nu} \partial^\rho F_{\rho\nu}; &
  & \cfgi{10} 
   \int \d \left ( \phi^\dagger \phi
  - \frac{v^2}{2} \right )
F_{\mu\nu}^2.
   \  \label{g.invs}
\end{align}

The coefficients are related to one-loop amplitudes according to the algebraic relations derived in~\cite{Binosi:2019nwz}.
Notice that they still hold in the presence of the deformation controlled by $z$ since they arise from the projection of the linearized ST operator, that is unaffected by the gauge-invariant term $\sim z \partial^\mu X_2 \partial_\mu X_2$.
We find in the cohomologically trivial sector
\begin{align}
 \cfct{1} = - \frac{1}{ev}  \overline{\Gamma}^{(1)}_{\chi^* \omega} = \frac{1}{8 \pi^2 v^2 }\frac{M_A^2}{1+z}\frac{\delta_{\xi;1}}{\epsilon} 
 ;
\end{align} 
\begin{subequations}
\begin{align}
	- m^2 v \cfct{0} + v \cfgi{1} &=\overline{\Gamma}^{(1)}_{\sigma};\label{tadsa}\\
	\cfct{0} v^2 + 
	%\cfxt{\bar c^*}
	\cfps{1}
	&=\overline{\Gamma}^{(1)}_{\bar c^*}.
	\label{tadsb}
\end{align}
\end{subequations}
In Feynman gauge one can safely set $\cfct{0}=0$
while $\cfct{1}=0$ in Landau gauge~\cite{Binosi:2019nwz} since there are no radiative corrections to the antifield-dependent amplitudes.
Hence we obtain ($\cfgi{1}$ is gauge-invariant):
\begin{align}
     \cfgi{1} &= \frac{1}{16 \pi^2 v^2}
    \left [
      ( m^2 + 6 M_A^2 ) M_A^2 +
      m^2 \frac{M^2}{(1+z)^2}   + 
      \frac{2 M^4}{(1+z)^3}  \right ]
      \frac{1}{\epsilon}  ; \nonumber \\
 \cfct{0} &= \frac{1}{m^2 v} \left ( \left . 
\overline{\G}^{(1)}_{\sigma} \right |_{\xi=1} -  \left . 
\overline{\G}^{(1)}_{\sigma} \right |_{\xi= 0} \right ) =  \frac{M_A^2}{16 \pi^2 v^2} \frac{\delta_{\xi ; 0}}{\epsilon}   .
\end{align}
By using the second of Eqs.(\ref{tadsb}) 
we get
\begin{align}
    \cfps{1} = - \frac{1}{16 \pi^2}
    \left [ M_A^2 + \frac{M^2}{(1+z)^2} \right ]
    \frac{1}{\epsilon}  ;
\end{align}
\begin{align}
      \cfps{3} &= %\cfxt{(\bar c^*)^2} =
        \left . \overline{\G}^{(1)}_{\bar c^*_1 \bar c^*_2} \right |_{p_1^2=0} = 
        \frac{1}{16 \pi^2} 
        \left [ 1 + \frac{1}{(1+z)^2} \right ]
        \frac{1}{\epsilon};&
      \cfps{5} &= %\cfxt{(\bar c^*)^2} =
        \left . \overline{\G}^{(1)}_{\bar c^*_1 \bar c^*_2 \bar c^*_3} 
        \right |_{p_2=p_3=0} = 0.  
\end{align}

Next, writing
\begin{align}
& \overline{\G}^{(1)}_{\bar c^*_3 \chi_1 \chi_2} = 
\ff{0(1)}{1;\bar c^*_3 \chi_1 \chi_2}+
\ff{1(1)}{1;\bar c^*_3 \chi_1 \chi_2}(p_1^2+p_2^2)+
\ff{2(1)}{1;\bar c^*_3 \chi_1 \chi_2}(p_1{\cdot} p_2),
\end{align}
we obtain
\begin{align}
&    \cfxt{1} = 
   \ff{0(1)}{1;\bar c^*_3 \chi_1 \chi_2} - 2 \cfct{0}
    -  2 \cfct{1}; \qquad
    \cfxt{3} = -\ff{2(1)}{1;\bar c^*_3 \chi_1 \chi_2} ;
    \qquad
    \cfxt{5} = -\ff{1(1)}{1;\bar c^*_3 \chi_1 \chi_2}
     .
\end{align}
An explicit calculation yields
\begin{align}
    \cfxt{1} = -\frac{1}{16 \pi^2 v^2}
    \left [ \underbrace{2 M_A^2 + m^2}_{\mathrm{0-sector}} -  \underbrace{\frac{2 M^2}{(1+z)^2}}_{\mathrm{1-sector}} +
   \underbrace{ \frac{m^2}{(1+z)^2} +
    \frac{4 M^2}{(1+z)^3}}_{\mathrm{2-sector}} \right ] \frac{1}{\epsilon},
    \label{cfxt.1}
\end{align}
where, as indicated, the first two terms in the square brackets arise from the St\"uckelberg subdiagrams (zero internal $X_2$-lines), the third from the subdiagrams with one internal $X_2$-lines and the last two terms from subdiagrams with $2$ internal $X_2$-lines. Notice that they are obtained from the corresponding expressions at $z=0$ by carrying out the replacement  $M^2 \rightarrow \frac{M^2}{(1+z)}$ and  multiplying each subdiagram by the appropriate prefactor $1/(1+z)^\ell$. Finally observe that $\cfxt{1}$ in  does not depend on the gauge, as it should being the coefficient of a gauge invariant operator.

Similarly we find
\begin{align}
    \cfxt{3} &= -\frac{1}{8\pi^2 v^2}
    \left [ 1 - \frac{1}{1+z} \right ]\frac{1}{\epsilon};&
    \cfxt{5}  &= 0.
\end{align}
Now we can fix $\cfxt{7}$
according to
\begin{align}
    & 2(\cfct{0} + \cfct{1}) + \cfxt{1}
    +  2 v^2 \cfxt{7} = \left . 
    \overline{\G}^{(1)}_{\bar c^*_3 \sigma_1\sigma_2} \right |_{p_1=p_2=0},
\end{align}
obtaining
\begin{align}
    \cfxt{7} = \frac{1}{4 \pi^2 v^4}
    \left [\underbrace{-\frac{M^2}{(1+z)^2}}_{\mathrm{1-sector}} + \underbrace{\frac{m^2}{(1+z)^2}+ \frac{4M^2}{(1+z)^3}}_{\mathrm{2-sector}} \underbrace{-\frac{m^2}{(1+z)^3}- \frac{3M^2}{(1+z)^4}}_{\mathrm{3-sector}} \right 
    ] \frac{1}{\epsilon},
\end{align}
where again we have collected factors of $1/(1+z)$ in such a way to identify their origin from the subdiagrams with a given number of internal $X_2$-legs. Notice that $\cfxt{3},\cfxt{7}$ vanish for $z \rightarrow 0$, in agreement with Ref.~\cite{Binosi:2019nwz} when the coupling constant $g$ of the operator $\sim\phi^\dagger \phi (D^\mu \phi)^\dagger D_\mu \phi$ is set to zero.

The amplitude 
\begin{align}
	\overline{\G}^{(1)}_{\bar c^*_3 A_1^\mu A_2^\nu} = \frac{M_A^2}{8 \pi^2 v^2}\frac{z}{1+z} \frac{g^{\mu\nu}}{\epsilon}
\end{align}
is momentum-independent both in Landau and Feynman gauge, which implies $\cfxt{9}=0$; in addition
\begin{align}
    \cfxt{13}= \frac{1}{2v} 
    \overline{\G}^{(1)}_{\sigma_3\bar c^*_1\bar c^*_2 } = \frac{1}{16 \pi^2 v^2}
    \frac{1}{(1+z)^2} \left (-1 + \frac{1}{1+z} \right ).
\end{align}
For $z \rightarrow 0$ one gets $\cfxt{13}= 0$,
again in agreement with Ref.~\cite{Binosi:2019nwz}.

We now move to the sector of gauge-invariant operators depending on the fields only. The coefficients of the potential $\cfgi{2}$ and $\cfgi{3}$ are derived by solving the equations
\begin{subequations}
\begin{align}
& 2 v^2\cfgi{2}+\cfgi{1}-2 m^2 \cfct{1} - 5 m^2\cfct{0} =
\left . 
\overline{\G}^{(1)}_{\sigma_1\sigma_2}
\right |_{p=0} ,\\
& 6 v^3 \cfgi{3} + 6 v \cfgi{2}
- \frac{9 m^2}{v} \cfct{1} -  \frac{12 m^2}{v} \cfct{0} = \left . 
\overline{\G}^{(1)}_{\sigma_1\sigma_2\sigma_3}
\right |_{p_2=p_3=0}.
\end{align}
\end{subequations}
It is convenient to express the results for each $\overline{\lambda}_{2,3}$ as the sum over the $\ell$-sector contributions; writing
\begin{align}
	\overline{\lambda}_{2,3} = \sum_{\ell=0}^2 \overline{\lambda}_{2,3}^{(\ell)}
\end{align}%
we obtain
\begin{align}
    & \cfgi{2}^{(0)} = \frac{1}{32 \pi^2 v^4}\frac{m^4 + 12 M_A^4}{\epsilon}; \qquad
    \cfgi{2}^{(1)} = \frac{1}{8 \pi^2 v^4}
    \left [ \frac{m^2 M_A^2}{1+z} -
    \frac{m^2 M^2}{(1+z)^2}  -\frac{2M^4}{(1+z)^3} \right ] \frac{1}{\epsilon}, &
    \nonumber \\
    & \cfgi{2}^{(2)} = \frac{1}{32 \pi^2 v^4}
    \left [ \frac{m^4}{(1+z)^2}  + \frac{8 m^2 M^2}{(1+z)^3} + \frac{12 M^4}{(1+z)^4}\right ] \frac{1}{\epsilon}   ,&
    \nonumber \\
    & \cfgi{3}^{(0)} = -\frac{m^2 M_A^2}{16 \pi^2 v^6}\frac{1}{\epsilon}  , \qquad
    \cfgi{3}^{(1)} = \frac{1}{16 \pi^2 v^6}
    \left [ 
    \frac{m^2 M_A^2}{1+z} + 
    \frac{4 m^2 M^2}{(1+z)^2} +
    \frac{8 M^4}{(1+z)^3}
    \right ]
    \frac{1}{\epsilon}  , \nonumber \\
    & \cfgi{3}^{(2)} = -\frac{1}{8 \pi^2 v^6}
    \left [ \frac{m^4}{(1+z)^2} + \frac{8 m^2 M^2}{(1+z)^3}+ 
    \frac{12 M^4}{(1+z)^4} \right ]
    \frac{1}{\epsilon}  , 
    \nonumber \\
    & \cfgi{3}^{(3)} = \frac{1}{8 \pi^2 v^6}
     \left [ \frac{m^4}{(1+z)^3} + \frac{6 m^2 M^2}{(1+z)^4}+ 
    \frac{8 M^4}{(1+z)^5} \right ]
    \frac{1}{\epsilon}  .
\end{align}
Notice that in the limit $z \rightarrow 0$ one recovers
the results given in~\cite{Binosi:2019nwz}, in particular
$\lambda_3$ becomes zero; furthermore, the consistency conditions (the 1-PI amplitudes in the right-hand side are understood at zero external momenta):
\begin{align}
& 2 v \cfgi{2} - \frac{3 m^2}{v} \cfct{1} - \frac{4 m^2}{v} \cfct{0} = \overline{\G}^{(1)}_{\sigma \chi\chi},
\nonumber \\
& 2 \cfgi{2} - \frac{4 m^2}{v^2} \cfct{1}
- \frac{4 m^2}{v^2} \cfct{0}
=
\overline{\G}^{(1)}_{\sigma\sigma \chi\chi},
\end{align}
hold separately for each $\ell$-sector.

The coefficients $\cfgi{4}$ and $\cfgi{5}$ are obtained from the relations
\begin{align}
& 2(\cfct{0}+\cfct{1})+\cfgi{4}
= \left . \frac{\partial\overline{\G}^{(1)}_{\chi_1\chi_2}}{\partial p^2} \right |_{p^2=0};&
&3 \cfgi{5} = \left . \frac{\partial\overline{\G}^{(1)}_{\chi_1\chi_2}}{\partial (p^2)^2} \right |_{p^2=0}.
\end{align}
The 2-point amplitude $\G^{(1)}_{\chi_1\chi_2}$ has UV degree of divergence $2$, hence $\cfgi{5}=0$; moreover
\begin{align}
    \cfgi{4}= -\frac{M_A^2}{8 \pi^2 v^2}\left ( 1 + \frac{3}{1+z} \right ) \frac{1}{\epsilon}.
\end{align}
In the limit $z\rightarrow 0$ we recover the results
of Ref.~\cite{Binosi:2019nwz} again by setting $g=0$.

The coefficients $\cfgi{6}$ and $\cfgi{7}$ can be determined from the 3-point function
\begin{align}
\overline{\G}^{(1)}_{\sigma_3 \chi_1 \chi_2} =
 \gamma^{0(1)}_{1;\sigma_3 \chi_1 \chi_2}+
 \gamma^{1(1)}_{1;\sigma_3 \chi_1 \chi_2} (p_1^2 + p_2^2)
 +  \gamma^{2,(1)}_{1;\sigma_3 \chi_1 \chi_2} p_1{\cdot} p_2 + {\cal O}(p^4)
\end{align}
where the momentum of the $\sigma$-field has been eliminated in favour of $p_{1,2}$ by imposing momentum conservation. Then $\cfgi{6},\cfgi{7}$ can be determined according to
\begin{align}
    &  2 v\cfgi{6}+ \gamma^{1(1)}_{1;\sigma_3 \chi_1 \chi_2} = 0;&
     2 v \cfgi{6}  + v \cfgi{7}
   + \gamma^{2;(1)}_{1;\sigma_3 \chi_1 \chi_2} = 0  .
\end{align}
Decomposing as before $\cfgi{6,7}$ according to the grading in the internal $X_2$-lines
\begin{align}
    \cfgi{6,7} = \sum_{\ell=0}^2 \cfgi{6,7}^{(\ell)}
\end{align}
one gets
\begin{align}
& \cfgi{6}^{(0)} = 0; 
\quad 
\cfgi{6}^{(1)} = \frac{1}{16 \pi^2 v^4}
\left [ 
\frac{2 M_A^2 }{1+z} +  \frac{ M^2}{(1+z)^2}
\right ] \frac{1}{\epsilon} ,
\nonumber \\
& 
\cfgi{6}^{(2)} = -\frac{1}{16 \pi^2 v^4}
\left [ \frac{2 M_A^2}{(1+z)^2} + \frac{M^2}{(1+z)^3} \right ] ,
\nonumber \\
&  \cfgi{7}^{(0)} = \frac{1}{8 \pi^2 v^4} \frac{2 M_A^2- m^2}{\epsilon}; 
\quad
\cfgi{7}^{(1)} = \frac{1}{8 \pi^2 v^4} 
\left [ 
\frac{2 M_A^2+m^2}{1+z} + \frac{M^2}{(1+z)^2}
\right ] \frac{1}{\epsilon}  , \nonumber \\
& 
\cfgi{7}^{(2)} = -\frac{1}{8 \pi^2 v^4}
\left [
\frac{4 M_A^2}{(1+z)^2} + \frac{M^2}{(1+z)^3} 
\right ] \frac{1}{\epsilon}  .
\end{align}
Again we notice that in the limit $z\rightarrow 0$
$\overline{\lambda}_{6,7}$ are zero, in agreement with the results
of Ref.~\cite{Binosi:2019nwz}.

There are no contribution of order $p^4$ in $\overline{\G}^{(1)}_{A_1^\mu A_@^\mu}$, so $\cfgi{9}=0$. In turn, $\cfgi{8}$ can be fixed by the projection equation
\begin{align}
     \left [   e^2 v^2 ( 2 \cfct{0} +  \cfgi{4}) + ( 2 \cfgi{8} + e^2 v^2
     \cfgi{5}) p_1^2  \right ] g^{\mu\nu} 
     + 2 \left ( e^2 v^2 \cfgi{5} - \cfgi{8}  \right ) p_1^\mu p_1^\nu 
     = \overline{\G}^{(1)}_{A_1^\mu A_2^\nu},
\end{align}
yielding
\begin{align}
    \cfgi{8} = - \frac{M_A^2}{48 \pi^2 v^2 (1+z)}\frac{1}{\epsilon}.
\end{align}
The limit $z \rightarrow 0$ reproduces the 
result of Ref.~\cite{Binosi:2019nwz}.

Finally, the coefficient $\cfgi{10}$ is recovered from the
amplitude $\overline{\G}^{(1)}_{\sigma_3 A_1^\mu A_2^\nu}$ 
\begin{align}
    \overline{\G}^{(1)}_{\sigma_3 A_1^\mu A_2^\nu}&=  \left[ \gamma^{0(1)}_{1;\sigma_3 A_1^\mu A_2^\nu}  -2 \gamma^{1(1)}_{1;\sigma_3 A_1^\mu A_2^\nu} p_1\cdot p_2 +
    \gamma^{2(1)}_{1;\sigma_3 A_1^\mu A_2^\nu} (p_1^2+p_2^2)\right] g^{\mu\nu}\nonumber \\
    & + \gamma^{3(1)}_{1;\sigma_3 A_1^\mu A_2^\nu} 
    p_1^\mu p_2^\nu + 
    \gamma^{4(1)}_{1;\sigma_3 A_1^\mu A_2^\nu} 
    p_1^\nu p_2^\mu,
\end{align}
as
\begin{align}
\cfgi{10} = \frac{\gamma^{1(1)}_{1;\sigma_3 A_1^\mu A_2^\nu}}{4v} = 0  .
\end{align}

%\bibliography{bibliography_1loop}


\end{document}